\documentclass[lettersize,journal]{IEEEtran}
\usepackage{cite}
\usepackage{amsmath,amssymb,amsfonts}
\usepackage{graphicx}
\usepackage{textcomp}
\usepackage{xcolor}
\usepackage{subcaption}
\usepackage{graphicx,epstopdf,psfrag,url,cite,xcolor,multirow,float}
\usepackage[font=footnotesize]{subfig}

\usepackage{algorithm}
\usepackage{algpseudocode}
\usepackage{array}
\usepackage{url}
\usepackage{verbatim}
\usepackage{diagbox}
\usepackage{threeparttable}

\usepackage[T1]{fontenc}
\usepackage{booktabs}

\begin{document}

\title{CollaPipe: Adaptive Segment-Optimized Pipeline Parallelism for Collaborative LLM Training in Heterogeneous Edge Networks}

\author{Jiewei Chen, Xiumei Deng, Zehui Xiong, Shaoyong Guo, Xuesong Qiu, Ping Wang, Dusit Niyato

%\author{Jiewei Chen,  ~\IEEEmembership{Graduate Student Member,~IEEE}, Shaoyong Guo, Xuesong Qiu,~\IEEEmembership{Senior Member,~IEEE}
%\author{IEEE Publication Technology,~\IEEEmembership{Staff,~IEEE,}
%        % <-this % stops a space
%\thanks{This paper was supported by the National Natural Science Foundation of China (62322103). \textit{(Corresponding author: Shaoyong Guo)}}
\thanks{Jiewei Chen, Shaoyong Guo, and Xuesong Qiu are with the State Key Laboratory of Networking and Switching Technology, Beijing University of Posts and Telecommunications, Beijing, China (e-mail: \{chenjiewei, syguo, xsqiu\}@bupt.edu.cn). 
Xiumei Deng is with the Singapore University of Technology and Design, Singapore (e-mail: xiumei\_deng@sutd.edu.sg).
Zehui Xiong is with the School of Electronics, Electrical Engineering and Computer Science, Queen's University Belfast, United Kingdom (e-mail: z.xiong@qub.ac.uk).
Ping Wang is with Dept. of Electrical Engineering \& Computer Science, Lassonde School of Engineering, York University, Canada (e-mail: ping.wang@lassonde.yorku.ca).
Dusit Niyato is with College of Computing and Data Science, Nanyang Technological University, Singapore (e-mail: dniyato@ntu.edu.sg).}

}

\markboth{}%
{Shell \MakeLowercase{\textit{et al.}}: A Sample Article Using IEEEtran.cls for IEEE Journals}


\maketitle

\begin{abstract}
The increasing demand for intelligent mobile applications has made multi-agent collaboration with Transformer-based large language models (LLMs) essential in mobile edge computing (MEC) networks. However, training LLMs in such environments remains challenging due to heavy computation, high end-to-end latency, and limited model generalization.
We introduce CollaPipe, a hybrid distributed learning framework that integrates collaborative pipeline parallelism with federated aggregation to support self-evolving intelligent networks. In CollaPipe, the encoder part is adaptively partitioned into variable-sized segments and deployed across mobile devices for pipeline-parallel training, while the decoder is deployed on edge servers to handle generative tasks. Then we perform global model update via federated aggregation.
To enhance training efficiency, we formulate a joint optimization problem that adaptively allocates model segments, micro-batches, bandwidth, and transmission power. We derive and use a closed-form convergence bound to design an Dynamic Segment Scheduling and Resource Allocation (DSSDA) algorithm based on Lyapunov optimization, ensuring system stability under long-term constraints.
Extensive experiments on downstream tasks with Transformer and BERT models show that CollaPipe improves computation efficiency by up to 15.09\%, reduces end-to-end latency by at least 48.98\%, and cuts single device memory usage by more than half, enabling online learning in heterogeneous and dynamic communication environments.
\end{abstract}


\section{Introduction}
With the rapid development of artificial intelligence generated content (AIGC) technologies in mobile Internet of Things (IoT), AI agent systems powered by large language models (LLMs) are emerging as a critical enabler for next-generation intelligent applications in mobile edge computing (MEC) networks \cite{wang2024FT-AIGC,guan2024citygpturbaniotlearning,kok2024IoTLLM}. 
By leveraging the strong generalization and content generation capabilities of LLMs, such systems can facilitate autonomous learning and decision-making across distributed IoT devices, supporting diverse scenarios such as UAV-based inspection in substations \cite{liu2025agent-inspection} and intelligent maintenance with industrial robots \cite{wang2024GenerativePFM}. In the future, MEC networks are expected to accelerate the deployment of distributed LLM services by providing massive data, idle computing resource, and native support for autonomous collaboration and continual learning across heterogeneous networks \cite{xu2025integratedLearnCom,li2024surveyLLMmultiagent}.

However, optimizing LLMs typically requires access to sensitive user data, raising privacy concerns and limiting adoption in many domains. To address this, recent research has explored the use of LLM-based federated learning (FL) to enable collaborative training and inference without centralized data sharing \cite{yu2024federatedFM,wu2024LLMtuningFL,wang2025RobustFLforLLMinAdvWireless}. By integrating distributed knowledge through local model training, FL paradigm mitigates the risk of user data leakage while enhancing the generalization of edge models by embedding global knowledge \cite{joshua2024FLPrivacySurvey, guangxi2023GradientAttacksFL,meng2024LocalPersonalization}.
Despite these benefits, the FL framework faces significant challenges in practical deployment for LLM-driven IoT applications. First, the substantial computational and memory demands of LLMs often exceed the capacity of individual devices. Second, collaborative distributed training of LLMs entails frequent communication, where the FL paradigm suffers from excessive communication overhead and inefficient resource utilization \cite{hu2025FLLM}.

Recent efforts have aimed to improve the computational and communication efficiency of LLM training through split learning frameworks. For instance, Wang \textit{et al.} \cite{wang2024federated_fine-tuning} proposed a split federated learning approach for fine-tuning pre-trained models, where the computationally intensive encoder is placed on the edge server, while the embedding and task-specific modules are deployed on edge devices. While this approach reduces the computational burden on devices, it shifts most of the workload to the edge server and lacks a collaboration mechanism for constructing a global model across distributed nodes. As a result, it fails to support cross-domain collaborative learning, particularly in IoT scenarios where data remains siloed across diverse domains.

Existing distributed intelligence paradigms also face limitations in accommodating the architectural complexity of Transformer-based LLMs. Traditional split learning and partitioning approaches typically focus on identifying an optimal cut layer between device-side and server-side models \cite{junhe2025FSL_pruning}. However, this coarse-grained partitioning often fails to fully utilize available network and computational resources, resulting in limited learning performance and system efficiency \cite{zhang2020NetworkBottleneck}. Conversely, overly fine-grained segmentation of LLMs increases the frequency of inter-device communication, introducing significant overhead and potential bottlenecks. 
Implementing adaptive LLM training in heterogeneous networks presents a critical challenge. 
Zheng \textit{et al.} \cite{zheng2024Split6G} demonstrated a tailored 6G architecture and a resource management scheme to support split edge learning for LLM training on mobile devices.
However, key factors, such as the interplay between communication overhead and resource allocation strategy, as well as the impact of device participation scale on overall learning performance for the LLM, remain insufficiently explored \cite{liu2025edgecloudcollaborativecomputing}.
Most existing wireless learning frameworks treat communication and training as separate processes, without considering how the collaborative learning paradigm influences both model performance and overall system efficiency \cite{dinh2021ConvergenceAndResourceAllocation, shi2021JointDSRA, shen2023JointOptimizationUAV, kim2024edge-cloud_controlled_lossy}. 
In addition, current collaborative intelligence paradigms primarily concentrate on split learning for deep neural networks (DNNs) \cite{wang2023ACE, xiumei2023LowLantencyFL_DNNPartition, hu2024CoLLaRS}, while offering limited exploration of quantitative evaluations for training Transformer-based LLMs and enabling generative AI tasks.
Therefore, integrating learning performance and communication efficiency into a \textit{bidirectional optimization framework}, achieving a dynamic balance through joint scheduling of resources and learning strategies, has become a critical direction for distributed collaborative intelligence in IoT.

To address these challenges, we propose \textit{CollaPipe}, a hybrid distributed learning framework integrates pipeline parallelism and federated aggregation. CollaPipe incorporates key system parameters, including the number of LLM segments, participating devices, and transmission power, into a unified optimization framework to support collaborative LLM training. Furthermore, we design a long-term dynamic device scheduling and resource allocation strategy to minimize end-to-end system latency while ensuring convergence of hybrid parallel training. To the best of our knowledge, this work is the first to introduce an adaptive collaboration framework that unifies pipeline parallelism and FL within edge-device network architectures, supporting the autonomous and continuous evolution of mobile AI agents at the network edge.
Our key contributions are summarized as follows:

\begin{enumerate}
\item We develop a hybrid distributed learning framework (CollaPipe) over heterogeneous edge networks by decomposing the LLM into embedding, encoder, and decoder (for generative tasks) modules. Specifically, the embedding module is deployed on the control units (CUs) within each cluster, the decoder is hosted on the edge server, and the computationally intensive encoder is partitioned into variable-sized segments, which are adaptively deployed across heterogeneous devices within each cluster. 
These modules are sequentially connected via wireless links, enabling efficient forward and backward propagation across the network.  
By facilitating fine-grained coordination of data, model components, and computational resources while preserving data privacy, CollaPipe supports efficient cross-domain collaboration. Compared to existing distributed learning paradigms, it offers a more effective balance between generalization performance and resource efficiency under constrained wireless edge environments.

\item We propose an optimized pipeline parallelism mechanism that adaptively determines the number of micro-batches ($m$) and model segments ($S$), reduce training runtime in clustered edge environments. A closed-form expression is derived to quantify both the latency and energy consumption when $K$ devices cooperatively process $S$ segments using pipeline parallelism. Furthermore, we provide the first convergence analysis of CollaPipe. Our analysis highlights the necessity of consistent communication and computation times per segment and reveals how the number of encoder segments and wireless uplink interference affect overall system performance. Based on these insights, we formulate a convergence-aware online system latency minimization problem, which jointly optimizes encoder partitioning, intra-cluster device scheduling, inter-cluster bandwidth allocation, and power control, under a $T$-round convergence constraint.

\item To address the complexity of jointly optimizing device scheduling and resource allocation across communication rounds, we employ Lyapunov theory to decouple the problem into two alternating subproblems, each corresponding to a single round. System stability is evaluated through a convergence gap metric. Building on this formulation, we design a Dynamic Segment Scheduling and Resource Allocation (DSSDA) algorithm and provide a structural analysis of the subproblems to establish the theoretical optimality of the proposed online algorithm.

\item We conduct extensive experiments to validate the effectiveness of the proposed framework and optimization algorithm. Using both the Transformer and BERT models, we evaluate performance on the machine translation, named entity recognition and sentence classification tasks. 
Experimental results demonstrate that our scheme not only achieves competitive training and inference performance but also consistently outperforms conventional pipeline-parallel and FL approaches, with improvements of up to 2.76\% in inference accuracy, and 48.98\% in training efficiency.
\end{enumerate}

The rest of the paper is organized as follows. In Section \ref{s2}, we introduce the related work on Transformer-based LLM partition, pipeline parallelism and edge computing. Then, Section \ref{s3} provides the CollaPipe framework for the distributed training. In Section \ref{s4}, we analyze the convergence behavior and provide the problem formulation. Section \ref{s5} proposes problem transformation method and the DSSRA algorithm. Experiments and results are provided in Section \ref{s6}. Finally, Section \ref{s7} concludes this paper. 
Table \ref{etab1} lists the main notations used in this paper.

\section{Preliminaries And Related Work}\label{s2}
\subsection{Model Parallelism and Edge Computing}
In the field of parallel computing for large-scale models, most existing works focus on co-training a complete model across multiple accelerators (i.e. GPU) in the cloud data center \cite{narayanan2021LLM-GPU,jay2020HetPipe}. 
The parallelism technologies aim to enable terminal devices to train large-scale models efficiently and accelerate the learning process. In data parallelism (DP), each worker holds a complete model replica, and input data is partitioned across workers \cite{brendan2017CommunicationEfficient_Data}. However, this approach is unsuitable for large-scale models due to memory constraints. Hence, the model parallelism (MP) was introduced to address this limitation by partitioning the model itself and distributing its segments across different workers \cite{ericp2015DistributedML,yonghao2023OptCommunicationModelPrallel}. 
Pipeline parallelism (PP) is a specific instance of MP in which the model is partitioned into sequential stages to reduce memory usage on individual workers. This strategy minimizes idle time across workers and achieves efficient training of large models \cite{huang2019GPipe,weigang2023MixPipe,xupeng2023SDPipe}. 

Current training clusters commonly adopt hierarchical network architectures, which improve communication efficiency by fully utilizing the bandwidth of each network layer \cite{shengwei2024HybridParallelDNN}. However, implementing a cross-device hybrid parallel strategy in heterogeneous wireless edge networks presents greater challenges due to the resource-constrained nature of terminal devices and the unstable communication environment \cite{yoon2022EdgePipe,hongjian2023CustomizedCloudServicein6G}.
Deng \textit{et al.} developed a cross-device collaborative fashion design system that enables distributed inference across different devices with optimized computation offloading \cite{hanhui2024CrossGAI}. However, the system does not take into account the efficiency and quality of model training tasks. 
Liao \textit{et al.} also presented an efficient split federated learning (SFL) method for edge computing, which controls the local updating frequency and batch size to accelerate model training \cite{yunming2024FL_DataModelParallelism}. Nevertheless, this method fails to consider network resource constraints, limiting its applicability and scalability for SFL computation in edge networks.
Chen \textit{et al.} introduced a fault-tolerant pipeline-parallel distributed training approach for heterogeneous edge devices \cite{yuyao2024FTPipeHD}. This method can periodically evaluate devices' computing power and resource status, allowing for dynamic adjustment of the deep neural network (DNN) partitioning strategy. However, it does not address issues on device energy consumption and power control in heterogeneous networks. Moreover, since LLM encoder training involves repeatedly applying the same Transformer block for feature representation, designing a dynamic cut layer has limited impact on improving learning performance.

Therefore, in this paper, we adopt pipeline parallelism exclusively for LLM encoder training \cite{lei2024AdvancesPipelineModel}, where each worker adaptively executes a variable number of Transformer layers as an encoder segment. At edge server, data parallelism is applied across different clusters to support full LLM training. This hybrid-parallel approach enables efficient utilization of heterogeneous devices while maintaining scalability. For the communication protocol of partitioned segments in production environments, we follow the design principles established in related works \cite{jakub2020DNNGraphOperators,su2024titanic}.

\begin{table}[t!]
	\caption{Notation declarations \label{etab1}}
	\centering
	\begin{tabular}{ |>{\centering\arraybackslash}m{1.7cm}|>{\raggedright\arraybackslash}m{6cm}|}
		\hline
		{\bf{Symbols}} & {\bf{Description}} \\
		\hline
		$B_n^U$ & Bandwidth allocated for the uplink transmission \\
		\hline
		$B^{dd}$ & Bandwidth allocated for the D2D transmission \\
		\hline
		$\hat{b}$ & Micro-batch size \\
		\hline
		$\mathcal{D}_n$ & Dataset of the $n$-th cluster \\
		\hline 
		$E_n^{com}$ & Energy consumption of the $n$-th CU for transmitting the encoder parameter   \\
		\hline
		$E_k^{sch}$ & Energy consumption of the $k$-th device for D2D transmission \\
		\hline
		$E_n^{pipe}$ & Energy consumption for $n$-th encoder training \\
		\hline
		$F_n$&  Loss function of the $n$-th local model \\
		\hline 
		$g^s$ & Smash data's gradients calculated by the $s$-th segment during backward propagation  \\
		\hline 
		$h_n$& Channel gain between the CU and BS \\
		\hline
		$h_{dd}$& Channel gain between devices \\
		\hline
		$\boldsymbol{I}_{n,j}\in \{0,1\}$ & Channel assignment matrix  \\
		\hline
		$I_i, I_{dd}$  & Interference over channels   \\
		\hline
		$\mathcal{J}$  & Index set of available channels   \\
		\hline
		$\mathcal{K}$ & Set of IoT devices within a cluster \\ 
		\hline
		$L$ & Number of TEBs of a LLM encoder \\
		\hline
		$\boldsymbol{m}$ & Number of micro-batches for pipeline parallelism  \\
		\hline  
		$\mathcal{N}$ & Set of clusters (FL participants) \\
		\hline
		$o_l,o'_l$ & FLOPs of the forward and backward propagation in the one TEB for each pipeline, respectively \\
		\hline
		$\boldsymbol{p}_n$ & Transmit power of the $n$-th CU \\
		\hline
		$p_k$ & Transmit power of the $k$-th device within a cluster  \\
		\hline
		$\boldsymbol{S}$ & Number of encoder segments ($S\le K$)  \\
		\hline
		$t=\{1,\ldots,T\}$ & Communication rounds between the server and CUs (i.e., epochs of FL) \\
		\hline
		$z^s$ & Smash data from the $s$-th segment during forward propagation \\
		\hline 
		$\boldsymbol{\delta}_k$ & Number of encoder blocks assigned on device $k$  \\ 
		\hline
		$\theta_n^{s}$ & Parameters of the $s$-th segment in cluster $n$ \\
		\hline
		$\theta_n^{enc}$ & Parameters of the $n$-th encoder \\
		\hline
		$\widetilde{\theta}^{enc}$ & Parameters of the global encoder \\
		\hline
		$\tau_n^{U}$ & Uplink transmission delay of the CU $n$ \\
		\hline
		$\tau_n^{cop}$ & Computation delay for pipeline training within $n$-th cluster  \\
		\hline
		$\tau_k^{dd}$ & D2D transmission delay of device $k$  \\
		\hline
		$\tau_k^{\text{pipe}}$ & Cross-device pipeline training delay   \\
		\hline
		$\tau(t)$ & Total latency of the $t$-th communication round \\
		\hline
		$\phi_k$ & FLOPs per clock cycle of the $k$-th device. \\
		\hline
	\end{tabular}
\end{table}

\subsection{Resource Management for Training Services}
From the perspective of network resource management, the rapid growth of mobile devices and applications has led to edge servers nearing critical resource shortages \cite{anselme2023DQL_SegmentRouting}. As a result, efficient resource allocation strategies are essential to reduce network latency in LLM training and enhance user satisfaction. 

Liu \textit{et al.} \cite{fangzheng2023Subtask} demonstrated that when mobile devices, edge servers, and cloud servers collaborate effectively through improved situational awareness, the overall task processing performance can be significantly enhanced. The authors also provided closed-form expressions for task partitioning ratios and resource allocation strategies \cite{dinh2021ConvergenceAndResourceAllocation}. Although this work considers the parallel and sequential structures of computational subtasks \cite{mahmoodi2019SchedulingOffloading}, the characterization of training tasks, particularly LLM training, is inherently more complex in practice.

Similarly, as we propose a Transformer Encoder Block (TEB)-wise model partitioning strategy, which differs from convolutional layer-based DNN model training, the scheduling of LLM training tasks and the corresponding resource allocation in edge networks require further investigation, which is the focus of this paper.

\begin{figure*}[!t]
	\centering{\includegraphics[width=7in]{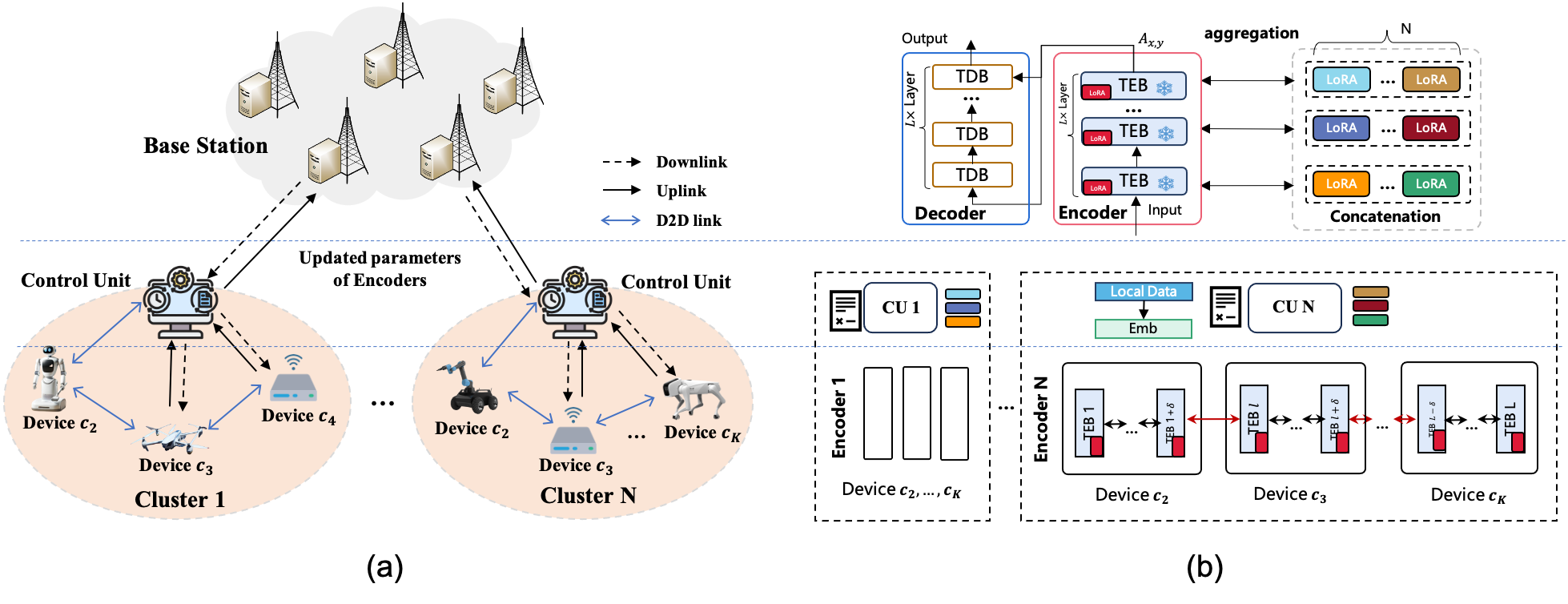}}
	\caption{Illustration of LLM module deployment over a distributed mobile edge IoT architecture. (a) A distributed mobile edge IoT scenario with device–device and device–edge collaboration; (b) Modularized LLM components mapped onto heterogeneous devices in (a). }
	\label{fig1}
\end{figure*}

\section{Hybrid Parallel Computing Framework in Heterogeneous Edge Networks}\label{s3}
In this section, we introduce the novel CollaPipe framework for collaborative LLM training, aiming to implement the efficient LLM services for mobile AI applications.

\subsection{System Model}\label{s3.1}
We consider a two-tier hierarchical network architecture comprising the edge server and multiple clusters \cite{xu2024Task-OrientedAIoT}, as illustrated in Fig.~\ref{fig1} (a). The edge servers consists of $N$ base stations (BSs), collectively forming an edge-cloud network. Within the network, devices are organized into clusters, each containing $K$ devices. A designated Control Unit (CU) serves as the cluster head, managing data storage and coordination within the cluster \cite{wang2023ACE, wei2025ComCluswithDP, liu2023EfficientD2DNet}.

The Transformer-based LLM is modularized and deployed as shown in Fig.~\ref{fig1} (b). The edge server is responsible for initializing the complete global model, training the decoder module, and aggregating local models. Within each cluster, pipeline parallelism is employed to pre-train the encoder module. The CU stores local training data and executes the embedding layer, while encoder segments are adaptively assigned to devices based on their capabilities. These devices then collaborate to train the encoder module in a distributed, parallel manner.

To enable efficient collaboration in heterogeneous networks, we propose CollaPipe, which organizes the learning process into two levels: device-to-device (D2D) collaboration within clusters and device-to-edge (D2E) collaboration across the network.

\subsubsection{D2D Collaboration}
As shown in Fig.~\ref{fig1} (a), each cluster consists of $K$ devices and one CU, interconnected through D2D communication \cite{liu2023EfficientD2DNet}. The LLM encoder is divided into $S$ sequential segments, with each segment comprising a set of Transformer Encoder Blocks (TEBs). The number of TEBs assigned to device $k$ is denoted by $\delta_k$, and the device set is represented by $\mathcal{K} = {1, 2, \ldots, K}$. During training, devices collaboratively execute pipeline-parallel learning by exchanging intermediate activations, labels, and gradients, enabling efficient distributed training across heterogeneous devices.

\subsubsection{D2E Collaboration}
The CUs from $N$ clusters are responsible for transmitting the local encoder parameters to the BS for FL, determining segment granularity and matching strategy for pipeline parallelism. 
In each communication round, the CU collects and concatenates the segments' parameters to assemble the complete encoder, and subsequently uploads the parameters to the BS for further integration and federated aggregation.

Let $\mathcal{J}=\{1,2,\ldots,J\}$ denote the index set of the available channels. Orthogonal frequency division multiplexing (OFDM) is utilized to transmit the local encoder parameters from large-scale factories to the BS in parallel. In each communication round, the BS determines a cluster-selection strategy, selects $J$ CUs and allocates available channels to them. After receiving the updated encoder parameters from different clusters, the BS conducts subsequent training of the decoder module and performs federated aggregation to update the parameters of the global LLM.

\begin{figure*}[t!]
	\centering
	\includegraphics[width=5 in]{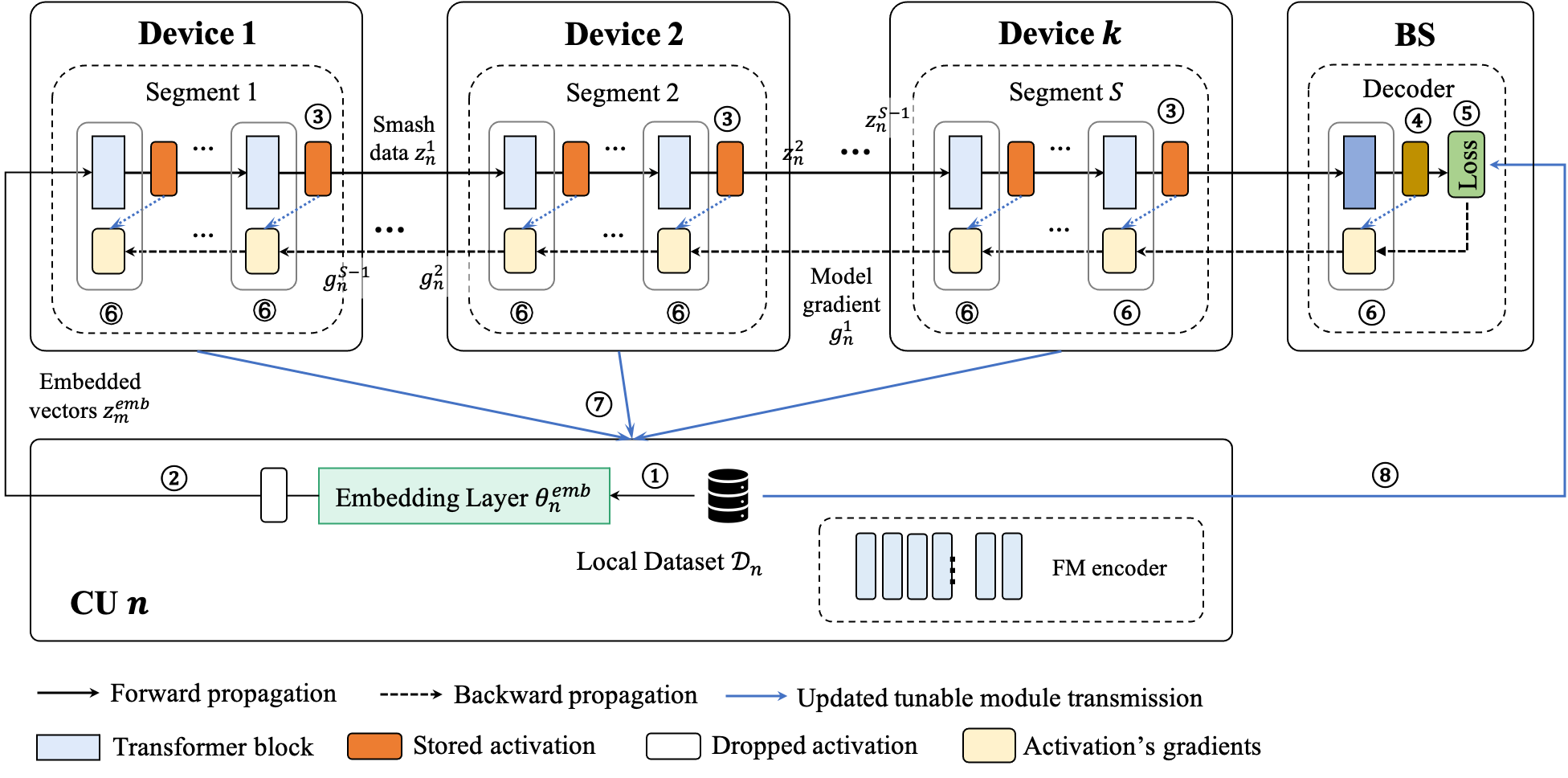}
	\caption{Workflow of CollaPipe with one iteration in heterogeneous networks.\\
		\textcircled{1}Data preprocessing; \textcircled{2}Execute embedding layer and ouput the embedded vectors; \textcircled{3}Execute partitioned segments and output the intermediate activations; \textcircled{4}Execute LLM decoder and compute the intermediate activations; \textcircled{5}Output the sequence/[CLS]sign, compute the loss between target and predicted values; \textcircled{6}Compute and output the gradients of loss with respect to the weight of partitioned segment or decoder, transmit the partial gradients to previous device, then update its own partitioned segment or decoder’s weights via the gradients. \textcircled{7}Transmit the updated segment weights to the CU; \textcircled{8}Transmit the updated encoder weights to the server.
	}
	\label{fig2}
\end{figure*}

\subsection{Learning Process}\label{s3.2}
We consider a LLM based on encoder-decoder architecture, with the Transformer backbone. The entire LLM is regard as consisting of two main modules, namely encoder and decoder. Inspired by \cite{du2024DistributedFM}, we partition the model into $S$ segments in a TEB-wise manner for collaborative deployment and training \cite{chen2025TB-wise}. 
The parameters of the $i$-th segment of cluster $n$ are parametrized by $\theta_n^i \in \mathbb{R}^p, i\in \mathcal{S}$, where $p$ represents the dimension of segment parameter vectors.
The encoder of the $n$-th cluster is parametrized by $\theta_n^{enc} \in \mathbb{R}^e$, where $e$ is the dimension of the parameter vector. 
Let $\mathcal{D}_n$ denote the initial local dataset of CU $n$. Each CU executes the embedding layer on $\mathcal{D}_n$ and it is responsible for selecting $S$ devices to execute the encoder sequentially. 

In practical mobile edge scenarios, the computing power and memory space of terminal devices vary significantly. Therefore, it is necessary to design efficient TEB-wise allocation, system power control, and wireless resource scheduling strategies to ensure the efficiency of the edge learning system and the quality of delivered models.
We assume that all devices within each cluster have different computational capabilities, while different clusters have one CU interacting with the BS. The BS typically possesses significant computing and communication resources. Inspired by the GPipe algorithm \cite{huang2019GPipe}, we design a hybrid parallel strategy and propose a learning algorithm of CollaPipe. 

The objective of CollaPipe is to identify the optimal LLM model parameters $\widetilde{\theta}^{\ast}$ that minimize the training loss. Let $F_n(\cdot)$ represent the local loss function for cluster $n \in \mathcal{N}$, then we have the following learning objective: 
\begin{equation}\label{argmin}
\begin{split}
	\widetilde{\theta}^{\ast} &= \arg \ \min_{\{\theta_n^{enc}\},\{\theta_n^s\}} G(\widetilde{\theta}^{FM};\theta^{enc}_n,(\mathbf{x},\mathbf{y})\in \mathcal{D}_n) \\
	&= \frac{1}{N}\sum_{n=1}^{N}\sum_{i=1}^{S} F_n \left( \theta_n^{FM}; \theta^i_n, (\mathbf{x},\mathbf{y})\in \mathcal{D}_n \right),
\end{split}
\end{equation}
where 
\begin{equation}\label{st-a}
	\theta_n^{FM} (t)= \widetilde{\theta}^{FM} (t),
\end{equation}
\begin{equation}\label{st-b}
	\theta_n^{FM} = \theta_n^{enc} \mathop{\oplus} \theta^{dec}, \forall n \in \mathcal{N},
\end{equation}
\begin{equation}\label{st-c}
	\theta_n^{enc} = \theta_n^1 \mathop{\oplus} \theta_n^2 \mathop{\oplus} \cdots \mathop{\oplus} \theta_n^S, \forall n \in \mathcal{N}.
\end{equation}
where $\theta^{FM}$ represents the parameters of the full model, $\theta^{dec}$ is the decoder parameters. In Eqs. (\ref{st-b}) and (\ref{st-c}), the symbol $\oplus$ denotes the concatenation operation. 

Specifically, in each learning epoch in one cluster, the following steps are sequentially performed.

\subsubsection{Key Hyperparameters Determining}
First, each cluster's CU executes embedding layer using its local data. The output from the embedding layer is then sent to the next terminal device. Consequently, the LLM encoder needs to be partitioned into several segments. These segments are executed sequentially from the CU to the last selected device $K$ using a micro-batch-based pipeline parallel algorithm. Therefore, the number of micro-batches, denoted as $m$, must also be determined. 
Let $b$ denote the overall batch size. The micro-batch size in pipeline parallelism is denoted by $\hat{b}$ and defined as:
\begin{equation}\label{unitdata}
	\widehat{b}=\frac{b}{m}.
\end{equation}

Since the pipeline-parallel learning process involves cross-device communication and scheduling, we require that each micro-batch's D2D transmission delay plus the computation delay be consistent, so that the working time of each stage will not overlap to ensure the effective operation of the algorithm. We will discuss the details in Section \ref{s3.4}.

\subsubsection{Segment Scheduling}
Since the computing, memory and radio resources of each terminal device are limited and their available idle resources fluctuate over time, the CUs are required to determine the number of TEBs to train on each device, ensuring efficient operation of the split-pipline-parallel learning.

\subsubsection{Local Forward Propagation of LLM Encoder}
Within each cluster $n$, forward propagation of the LLM encoder is executed during the $t$-th FL communication round using a pipeline scheduling approach. Each device $k$ assigned to the task receives intermediate activations $z^s$ from its preceding device $k-1$, performs local computations through multiple Transformer layers\footnote{We balance end-to-end latency through TEB-wise partition and segment scheduling, requiring only that forward and backward propagation follow the same path—without ordering segments or devices.}, generates new activations $z^{s+1}$, and passes them to the next device $k+1$.\footnote{In our designed strategy, not all devices are assigned to training tasks. $k$ is used to represent the task dependency between devices.}
The last device in the sequence uploads the activation of the encoder's final segment to the server, where the forward propagation of the decoder is performed.

\subsubsection{Global Training of LLM Decoder}
After receiving activations from $N$ clusters, the edge server at the base station directly feeds them into the decoder module. It computes the loss value based on the predicted output, calculates the gradient of the loss function with respect to the decoder's weights, and updates the model parameters of the decoder accordingly.

\subsubsection{Local Backward Propagation of LLM Encoder}
The last device in each cluster receives the decoder activation's gradients from the server. Subsequently, cluster $n$ performs backpropagation following the reverse order of the devices used during forward propagation, as Step 6 illustrated in Fig. \ref{fig2}.

\subsubsection{Global Model Aggregation And Updating}
After completing backpropagation, each segment and the decoder update their parameters using stochastic gradient descent (SGD) method. Each device sends its updated model parameters $\theta^s_n$ to its cluster's CU $n$, where the parameters are concatenated. The CU then uploads the encoder parameters $\theta^{enc}_n$ to the server, then FL aggregation is performed to update the global full model $\widetilde{\theta}^{FM}$ as follows:
\begin{equation}\label{eq6}
\widetilde{\theta}^{FM}(t+1) = \theta^{FM}(t)-\eta g^{FM}(t),
\end{equation}
where $g^{FM}(t)$ represents the aggregation gradients at the $t$-th round.

Following the chain rule, the sever calculates the averaged gradient of $N$ local full models as follows:
\begin{equation}\label{g-a}
	g^{FM}(t) = \frac{1}{N}\sum_{n=1}^{N} \nabla_{\theta_n} F_n(t),
\end{equation}
where
\begin{equation}\label{g-b}
	g_n^{FM} = g_n^{enc} \mathop{\oplus} g^{dec}, \forall n \in \mathcal{N},
\end{equation}
\begin{equation}\label{g-c}
	g_n^{enc} = g_n^1 \mathop{\oplus} g_n^2 \mathop{\oplus} \cdots \mathop{\oplus} g_n^S, \forall n \in \mathcal{N}.
\end{equation}

The process of CollaPipe is detailed in Algorithm 1.
Initially, pipeline parallelism is implemented within each cluster (Lines 5-23). Subsequently, FL is applied to facilitate cooperative training across clusters (Line 4 and Lines 24-28). 

\begin{algorithm}[!th]
	\caption{Hybrid Parallel Learning of CollaPipe}
	\begin{algorithmic}[1]
		\State \textbf{Input:} Number of clusters $N$; number of encoder segments $S$; number of TEBs $L$; Communication rounds (learning epochs) $T$; batch size $b$.
		\State \textbf{Output:} The optimal global LLM model $\widetilde{\theta}^{\ast}$.
		
		\State Initialize the full model $\theta^{FM}$;
		\For{each communication round $t=1,2,\ldots,T$}
		\State \underline{Forward Propagation of Encoder in Clusters:}
		\For{each cluster $n=1,2,\ldots,N$ in parallel}
		\For{each device from $1$ to $s$ within cluster $n$ in sequence}
		\State Perform forward propagation using pipeline parallelism with $m$ micro batches;
		\State Send the intermediate activations of each segment $z^s$ to the CU;
		\EndFor
		\State Send the intermediate activations of the encoder $z^{enc}$ to the server;
		\EndFor
		
		\State \underline{Decoder Traning on The Server:}
		\State Perform forward and backward propagation of the decoder from $N$ clusters;
		\State Calculate the gradients and send the intermediate gradients $g^{dec}$ to CUs;
		
		\State \underline{Backward Propagation of Encoder in Clusters:}
		\For{each cluster $n=1,2,\ldots,N$ in parallel}
		\For{device from $s$ to $1$ in sequence}
		\State Perform backward propagation of the encoder on each device through pipeline parallelism;
		\State Update the parameters of all segments $\theta_n^s$;
		\EndFor
		\EndFor
		\State Send segments' tunable parameters $\widehat{\theta}_n^s$ to the server for concatenation and aggregation;
		
		\State \underline{Global Aggregation on The Server:}
		\State Concatenate segments of the encoder and decoder, and perform global aggregation of the full LLM;
		\State Broadcast the updated encoder $\theta_{enc}$ to all clusters for next round training;
		\EndFor
		\State \textbf{Return} The updated full model $\widetilde{\theta}^{FM}(T)$
	\end{algorithmic}
\end{algorithm}

\subsection{Communication Model}\label{s3.3}
The CollaPipe framework operates within a heterogeneous network, where we consider a wireless edge system and a D2D communication scenario utilizing Orthogonal Frequency Division Multiple Access (OFDMA) technology.

\subsubsection{Device-to-Edge (D2E) Communication}
As for the D2E collaboration, the uplink rate of the CU from the $n$-th cluster is given by 
\begin{equation}\label{bandwidth2}
	r_n^{\text{Up}}=B_n^U \mathbb{E}_{h_n} \left(\log_2 \left(1+\frac{p_n h_n}{I_i+B_n^U N_0}\right)\right),
\end{equation}
where $B_n^U$ represents the allocated bandwidth for CU $n$ and $p_n$ is the transmit power of CU $n$. $h_n$ denotes the channel gain between CU $n$ and the BS. $\mathbb{E}_{h_n}$ is the expectation with respect to $h_n$ \cite{chen2021FL_over_wireless}. $N_0$ is the noise power spectral density, $I_i$ represents the interference caused by the CUs located far away from the service area \cite{aladin2024jamming}.

In this paper, we consider that the BS allocates sufficient bandwidth, so the downlink delay is ignored in the communication model construction.

Based on the transmission rate, in the $t$-th communication round, the transmission delays between the BS and each CU are derived as \cite{xiumei2023LowLantencyFL_DNNPartition}
\begin{equation}
\tau_n^{\text{Up}}(t) = \sum_{j=1}^{J}I_{n,j} \frac{z_n^{enc}+\theta_n^{enc}}{B_n^U(t)\mathbb{E}_{h_n} \left(\log_2 \left(1+\frac{p_n(t) h_n(t)}{I_i+B_n^U(t) N_0}\right)\right)},
\end{equation}
where $z_n^{enc}$ and $\theta_n^{enc}$ are the size of smash data and encoder model, respectively.

The energy consumption of the $n$-th CU for transmitting the concatenated encoder parameters in the $t$-th communication round is \cite{yang2023EnergySemantic}
\begin{equation}
	E^{\text{com}}_n(t)= p_n(t) \frac{\sum_{j=1}^{J}I_{n,j} \theta_n^{enc}}{B_n^U(t) \mathbb{E}_{h_n} \left(\log_2 \left(1+\frac{p_n(t) h_n(t)}{I_i+B_n^U(t) N_0}\right)\right)}.
\end{equation}

\subsubsection{Device-to-Device (D2D) Communication}
As for the D2D collaboration, our goal is to identify a sequence of devices $\widetilde{\mathcal{K}}(t)$ to instantiate the encoder segments. In the heterogeneous wireless communication system, devices communicate with each other using a D2D protocol within a cluster. For the sake of simplicity, we assume that the speed of D2D communication is steady, which is denoted as $r^{dd}$.

The data transmit delay from device $i$ to the subsequent device $j$ within a cluster can be calculated by the following formula:
\begin{equation}
\tau_k^{dd}(t)=\frac{z^s+g^{s+1}}{B^{dd} \log_2 \left(1+\frac{p_k(t) h_{dd}(t)}{I_{dd}+B^{dd}N_0}\right)},
\end{equation}
where $z^s$ represents the size of intermediate activations of the $s$-th segment during forward propagation, while $g^{s+1}$ is the smash data's gradients of the $(s+1)$-th segment during backward propagation. 

Accordingly, the energy consumption of the scheduled device $k$ in the $t$-th communication round is represented by
\begin{equation}
    E^{\text{sch}}_k(t) = p_k(t)\cdot \tau^{dd}_k(t).
\end{equation}

\begin{figure}[t!]
	\centering{\includegraphics[width=0.5 \textwidth]{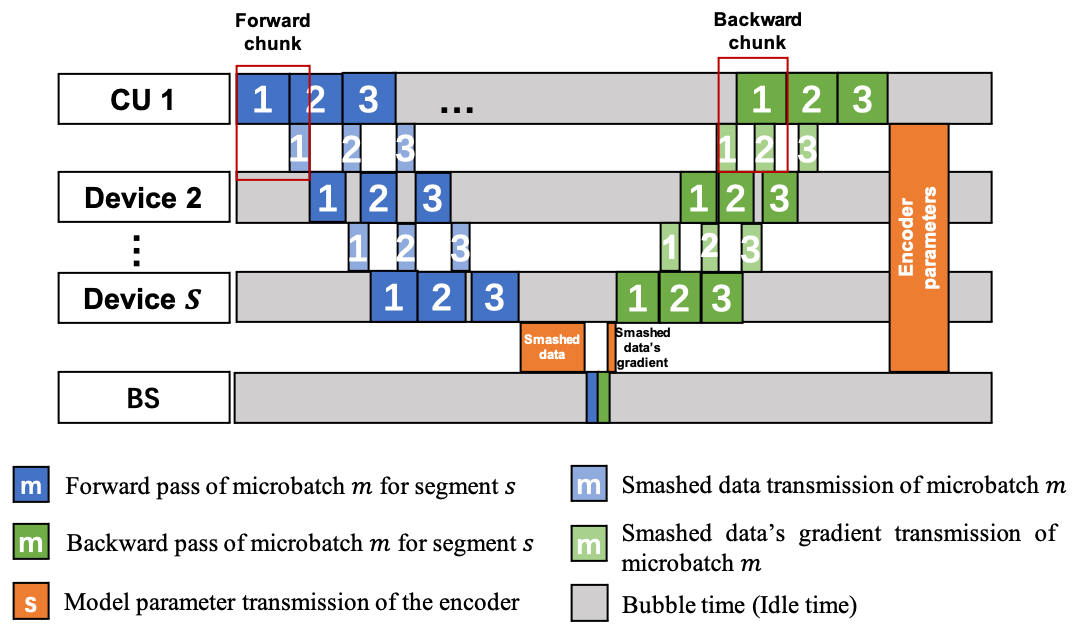}}
	\caption{Entire on-device training model with pipelines.}
	\label{fig3}
\end{figure}

\subsection{Pipeline Parallelism Model}\label{s3.4}
Considering the heterogeneity in devices' computational capabilities, each device is characterized by a relative computation factor, denoted as $f_k, \forall k$. 
In this paper, we apply the traditional pipeline parallel concept to achieve cross-device parallel training and reduce the communication overhead caused by multi-segment segmentation learning. As shown in Fig. \ref{fig3}, we consider the computation and communication overhead of a micro-batch as one forward chunk and one backward chunk, so that each micro-batch can perform pipeline calculations independently without clock overlap \cite{huang2019GPipe}. The entire training process under CollaPipe framework contains a total of $(S+m-1)$ chunks, where a chunk is the unit size of a micro-batch. Thus, the cross-device pipeline latency for the all scheduled devices $K$ in cluster $n$ during the encoder training at communication round $t$ can be expressed as 
\begin{equation}\label{device_cop}
\tau_{n,\delta_k}^{\text{pipe}}(t) = (S+m-1) \max_{k\in \mathcal{K}} \left( \frac{ \delta_k ( \hat{b} \cdot o_l+o'_l) }{\phi_k f_k} + \tau_k^{dd} \right)-\tau^{dd}_k,
\end{equation}
where $o_l$ and $o'_l$ represent the FLOPs of the forward and backward propagation for each pipeline in the $l$-th TEB, respectively; $\phi^D$ and $\phi^G$ represent the FLOPs per clock for forward and backward propagation for each pipeline in the $l$-th TEB, respectively \cite{xiumei2023LowLantencyFL_DNNPartition}. 

In Eq. (\ref{device_cop}), $\frac{ \delta_k ( \hat{b} \cdot o_l+o'_l) }{\phi_k f_k}$ represents the unit computation delay for training on the device, and $\tau_k^{dd}$ represents the unit delay for D2D transmission scheduling. Since the communication between the $s$-th device and BS is not in the pipeline delay model, the device scheduling time in one chunk is subtracted, i.e., $-\tau^{dd}_k$.

Within a cluster, the devices are responsible for executing the forward and backward operations of the matched segments. Therefore, the energy consumption of $K$ devices within the cluster $n$ for encoder training is given by
\begin{equation}
	E^{\text{pipe}}_n= 2m \left( \frac{\delta_k (\hat{b} \cdot o_l+o'_l)}{\phi_k} (f_k)^2 + E^{\text{sch}}_k \right).
\end{equation}

\section{Convergence Analysis and Problem Formulation}\label{s4}
In this section, we first derive a bound on model divergence to quantify the effects of model partitioning and signal transmission in heterogeneous network environments. Based on this analysis, we then formulate the corresponding optimization problem.

\subsection{Impact of Heterogeneous Network Environment and Model Splitting}\label{s4.1}
The segment number $S$ and micro-batch size $\hat{b}$ for collaborative LLM training can be determined based on the derived divergence bound in advance.
Furthermore, D2E collaboration involves long-distance communication. Therefore, the signal interference in the wireless environment needs to be considered. Therefore, our analysis of the model divergence bound focuses on three parts, i.e., the interference error, the number of segments $S$ and clusters $N$.

To facilitate the analysis, we consider several assumptions regarding the loss function, model weights, and gradients, as outlined below \cite{wang2024federated_fine-tuning,amiri2021ConverFLWireless,junhe2025FSL_pruning}.

\textbf{Assumption 1.} \textit{The loss function $F_n(\theta)$ is non-convex, differentiable and $\beta$-smooth, i.e.,}
\begin{equation}\label{ass1}
\Vert \nabla F_n(\theta_1) - \nabla F_n(\theta_2) \Vert \le \beta \Vert \theta_1 - \theta_2 \Vert, \forall \theta_1, \theta_2.
\end{equation}

\textbf{Assumption 2.} \textit{(Bounded gradient). The variance of all entries of local gradient $g^s_n$ is upper bounded by a constant $\phi$, i.e.,}
\begin{equation}\label{ass2}
	\mathbb{E} \Vert g^s_n \Vert^2 \le \phi^2.
\end{equation}

\textbf{Assumption 3.} \textit{There exists a constant $\xi \ge 0$ such that the Polyak-Lojasiewicz inequality holds for $F(\theta)$, i.e.,}
\begin{equation}\label{ass3}
\frac{1}{2}\Vert \nabla F(\theta) \Vert_2^2 \ge \xi \left(F(\theta)-F(\theta^{\ast})\right).
\end{equation}

The first three assumptions are commonly adopted in stochastic optimization and federated learning literature to facilitate convergence analysis under non-convex settings. Next, we provide the assumption for pipeline parallelism.

\textbf{Assumption 4.} \textit{The sum of computation and communication delays of all devices in pipeline parallel training tends to be consistent.}

Assumption 4 can be achieved by adjusting the model partitioning strategy (the number of TEBs $\delta_k$ in each segment and the size of a micro-batch $\hat{b}$, etc.) to avoid the overall pipeline being blocked due to some devices with excessive latency, which affects the convergence speed and performance of the model.

\textbf{Assumption 5.} \textit{(Stationary backpropagation). Expectation terms involving $\theta_n^s$ remain constant during training.}

Assumption 5 holds true for our TEB-wise split training: we can optimize the weights $\theta_n^s$ for a partitioned segment of the LLM's encoder while keeping the weights of other partitions fixed, thereby ensuring that the backpropagated statistics $g_n^s$ remain stationary.

\textbf{Lemma 1.} \textit{Given the objective in (\ref{argmin}), the convergence behavior of $F\left(\theta_n^{FM};\{\mathcal{D}_n \} \right)$ can be separately optimized with respect to the LLM's encoder $\{ \theta_n^{enc} \}$ and decoder (task module), i.e.,}
\begin{equation}\label{lemma1}
	\begin{split}
		\Vert \nabla F(\theta^{FM}) \Vert^2 &\le \underbrace{\frac{S^2}{N^2L} \sum_{n\in \mathcal{N}} \sum_{l=1}^{L/S}\sum_{s=1}^{S} \left\Vert \nabla_{\theta_n^{s,l}}F_n \right\Vert^2}_{\text{partitioned encoder module}} + \\
		&\quad \quad \quad \quad \underbrace{\frac{1}{N} \sum_{n\in \mathcal{N}} \left\Vert \nabla_{\theta^{dec}}F_n \right\Vert^2}_{\text{decoder module}},
	\end{split}
\end{equation}
\textit{where $L/S$ represents the average number of TEBs in each segment.}

Proof: See Appendix A.

We next propose a new lemma to characterize the impact of the communication environment on the convergence behavior.

\textbf{Lemma 2.} \textit{(Interference impact). Let $\theta_k, \hat{\theta}_k \in \mathbb{R}^E$ be the true and corrupted word embedding and $\mathbf{s}_{k}, \hat{\mathbf{s}}_{k} \in \mathcal{C}$ be the corresponding true and corrupted transmit symbols (signals). According to \cite{aladin2024jamming}, it holds that}
\begin{equation}
	\mathbb{E} \left[ \Vert F(\theta^{enc}_n)-F(\hat{\theta}^{enc}_n) \Vert^2 \right]  \le \beta\mathbb{E} \left[ \Vert \mathbf{s}_n - \hat{\mathbf{s}}_n \Vert^2 \right]. 
\end{equation}

In Lemma 2, each CU $n$ transmits the signal $x_n$ which is composite of the beamforming vector $w$ and word embedding symbols $s_n$. Then, the mean square error of signal transmit depends on the transmitted power $p_n$ of sender (CU) $n$, we have \cite{nahum2024RadioResource}
\begin{equation}
\mathbf{s}_n= PL \cdot SF \cdot \left(|l_{los}|^2+\sum_{n=1}^{N}\sum_{z=1}^{Z_{ray}}|l_{nlos}|^2\right)\cdot p_n,
\end{equation}
where $PL$ and $SF$ represent the path loss and shadow fading between the CU and BS, respectively. $l_{los}$ is the line-of-sight (LOS) path contribution, and $l_{nlos}$ is non-line-of-sight (NLOS) paths for each ray $z$ and cluster $n$, where $Z_{ray}$ is the number of rays \cite{3GPP20173Dchannel, nahum2024RadioResource}.

According to Lemmas 1, 2 and Assumptions 1 to 5, we have the following theorem.

\textbf{Theorem 1.} \textit{Given the arbitrary bandwidth allocation and device segment matching policy, when $\eta< \frac{4\xi NL}{\beta (S^2+L)}$, the optimality gap after $T$ communication rounds is upper bounded by}
\begin{equation}
	\begin{split}
		&\quad F\left(\theta^{FM}(T)\right) - F\left(\theta^{FM*}\right) \\
		&\le \underbrace{\left( \prod_{t=0}^{T-1}(1-2\sigma(t)) \right)F(\theta^{FM}(0))-F(\theta^{FM*})}_{\text{initial \ gap}} \\
		&\quad + \underbrace{\sum_{t=0}^{T-1}\prod_{j=t+1}^{T-1}\left(1-\sigma(j)\right)\frac{\beta \eta^2 \phi^2}{N}\left(\frac{S^2}{L}+1\right)}_{\text{task\ related\ gap}} \\
		& \quad + \underbrace{ \sum_{t=0}^{T-1}\prod_{j=t+1}^{T-1}\left(1-\sigma(j)\right)\frac{\eta}{N} \epsilon(p_n)}_{\text{interference\ related\ gap}},
	\end{split}
\end{equation}
\textit{where}
\begin{equation}\label{sigma}
	\sigma(t)=\eta\xi-\frac{\beta \eta^2}{2}\left(1+\frac{S^2}{NL}\right),\\
\end{equation}
\begin{equation}\label{epsilon}
	\epsilon(p_n) = \frac{C}{p_nh_n+I_i}.
\end{equation}

Proof: See Appendix B.

\textit{Remark 1: The granularity of parallelism has a dual effect on convergence behavior, and increasing the number of segments does not necessarily lead to faster convergence.}
The number of model segments $S$ reflects the granularity of pipeline parallelism and directly affects both the training latency and convergence behavior. As shown in Theorem~1, $S$ appears quadratically in the task-related term:\[\frac{\beta\eta^2 \phi^2}{N} \left( \frac{S^2}{L} + 1 \right),\]
which indicates that excessively large $S$ amplifies the local model divergence due to partial gradient computation in distributed clusters. Moreover, $S$ indirectly affects the per-round delay and thus limits the number of global updates $T$ within a time budget. 
While increasing $S$ enhances parallelism and improves device utilization, it may also introduce diminishing returns in convergence due to elevated communication overhead and inconsistencies among sub-models. These findings suggest that $S$ should not be optimized independently. Instead, it should be jointly tuned with other system-level parameters, such as the number of micro-batches, to achieve a balanced trade-off between parallel efficiency and convergence performance under the CollaPipe framework.

\textit{Remark 2: Communication interference causes cumulative delays in convergence.}
In the interference-related term, the transmission power $p_n$ of each cluster's CU directly affects the signal distortion during the encoder parameter upload process. Specifically, the main gap component
\[ \frac{\eta}{N} \cdot \epsilon(p_n) = \frac{\eta}{N} \cdot \frac{C}{p_n h_n + I_i} \]
reveals that increasing $p_n$ leads to a lower transmission-induced error $\epsilon(p_n)$, thereby improving the accuracy of federated aggregation and accelerating global convergence. Moreover, the function $\epsilon(p_n)$ is monotonically decreasing and convex with respect to $p_n$, indicating diminishing returns of increasing $p_n$ under high-power regimes.
This observation guides the design of the CU's power control module when designing the dynamic optimization algorithm. We expect that in each round of communication, higher power levels should be allocated to clusters with better channel gain $h_n$ or greater influence on the global model, so as to transmit local encoder parameters more accurately. 
At the same time, the energy constraint $E_n(t) = p_n(t) \cdot \tau_n(t)$ ensures that the improvement of convergence performance will not come at the cost of excessive energy consumption, so as to balance the convergence performance and resource efficiency in the wireless network.

\subsection{Problem Formulation}\label{s4.2}
According to the analysis above, the delay of each communication round is divided into two stages: encoder training via pipeline parallelism in each cluster; model parameter transmission under the edge FL framework. 
Thus, the total delay of the $t$-th communication round for LLM training is given by 
\begin{equation}
	\tau(t) = \max_{n\in \mathcal{N}} \ \{ \tau^{\text{pipe}}_{n,\delta_k}(t) +\ \tau^{\text{Up}}_n(t) \}.
\end{equation}

According to Theorem 1, the convergence of the CollaPipe is influenced by model partitioning, pipeline parallelism and power control strategies.
To obtain the communication-computation efficient LLM collaborative training framework, we develop a dynamic segment scheduling and resource allocation protocol. This protocol, guided by Theorem 1, explicitly incorporates model partitioning, pipeline parallelism and power control to minimize the average training delay while adhering to constraints on energy consumption and network resource usage. Let the scheduling protocol be $\boldsymbol{X}(t)=\left[\boldsymbol{m}, \boldsymbol{S}(t), \boldsymbol{\delta}_k(t), \boldsymbol{I}_{n,j}(t), \boldsymbol{p}_n(t) \right]$, we formulate a stochastic optimization problem as 
\begin{equation}\label{P0}
	\begin{split}
		\textbf{P0}: &\ \min_{\boldsymbol{X}(t)} \ \frac{1}{T}\sum_{t=1}^{T} \tau(t),\\
		\text{s.t.} &\quad \textbf{C1}: \sum_{k=1}^{K}\delta_k = L, \delta_k, m \in \mathbb{Z}^+,\\
		& \quad \textbf{C2}: S(t)=\sum_{k=1}^{K}1_{\{\delta_k>0\}}, 1\le S\le K, S \in \mathbb{Z}^+,\\
		& \quad \textbf{C3}: I_{n,j}(t) \in \{0,1\}, \forall n \in \mathcal{N}, j \in \mathcal{J},t\in \mathcal{T}, \\
		& \quad \textbf{C4}:\sum_{j\in \mathcal{J}} I_{n,j}(t) = 1, \forall n\in \mathcal{N}, t\in \mathcal{T},\\
		& \quad \textbf{C5}:\sum_{n\in \mathcal{N}} I_{n,j}(t) \le 1, \forall j \in \mathcal{J},t\in \mathcal{T},\\
		& \quad \textbf{C6}:0 \le p_n(t) \le P^{max}_n, \forall n \in \mathcal{N},t\in \mathcal{T},\\
		& \quad \textbf{C7}:0 \le \delta_k(t)\gamma_0 \le \gamma^{max}_k, \forall k \in \mathcal{K},t\in \mathcal{T},\\
		& \quad \textbf{C8}:0 \le E^{\text{com}}_n(t) \le E^{max}_n(t), \forall n \in \mathcal{N},t\in \mathcal{T},\\
		& \quad \textbf{C9}:0 \le E^{\text{pipe}}_k(t) \le E^{max}_k(t), \forall k \in \widetilde{\mathcal{K}},t\in \mathcal{T},\\
		& \quad \textbf{C10}:\frac{1}{T}\sum_{t=1}^{T} \Gamma \le \Gamma^{max}.
	\end{split}
\end{equation}
where $\gamma^k_{max}$ denotes the memory budget on device $k$, while $\gamma_0$ represents the memory required for one TEB training. $\Gamma$ is the balance bound of one-round training, according to the proof of Theorem 1, we have
\begin{equation}
\Gamma = \frac{\beta \eta^2}{2N} \left(\frac{\phi^2 }{L}S^2 +\frac{C}{p_nh_n+I_i} +\phi^2 \right).
\end{equation}

The ranges of the variables $\boldsymbol{m}, \boldsymbol{\delta}_k$ are constrained by \textbf{C1}, ensuring that all the TEBs can be trained by $K$ devices within one cluster; \textbf{C2} denotes that the number of encoder segments can not exceed the number of devices in a cluster; \textbf{C3}, \textbf{C4} and \textbf{C5} indicate that all the CUs cooperatively train the LLM via FL, and one local encoder will be transmitted to the server at each communication round; $p_n$ is constrained by \textbf{C6}; \textbf{C7} specifies that each device where a segment is deployed must ensure sufficient memory to fine-tune its corresponding segment; $\textbf{C8}$ and $\textbf{C9}$ are the energy consumptions for devices and CUs in each communication round, respectively; the long-term constraint \textbf{C10} is adopted to optimize the learning performance by balancing the number of segments and the CUs' power.

\section{Dynamic Segment Scheduling And Resource Allocation Algorithm}\label{s5}
In this section, we leverage Lyapunov optimization theory to transform the original problem \textbf{P0} into a per-round delay minimization problem subject to system stability constraints. To address the inherent non-convexity, we propose a dynamic optimization algorithm based on problem decoupling and an block coordinate descent method.

\subsection{Problem Transformation via Lyapunov Analysis}\label{s5.1}
Since the goal of our problem formulation is to optimize the average time, we transform the \textbf{P0} problem (\ref{P0}) into a virtual stable queue $Y_n(t)$ through Lyapunov analysis. The following formula is used to express the virtual queue $Y_n(t)$ of each CU $n$ related the delay and convergence constraint in a long-term state:
\begin{equation}\label{queue}
	Y_n(t+1)\triangleq \max \{Y_n(t)+\Gamma^t-\Gamma^{max}, 0 \},
\end{equation}
where $Y_n(t)$ is the virtual delay queue, denoting the cumulative degree of failure to meet convergence and delay requirements. 
$\Gamma^t$ represents the convergence bound at $t$-th communication round. $\Gamma^{max}$ is the maximum bound constraint. 

In order to ensure the optimality of the system in a long-term perspective, we add a stability constraint to the \textbf{P0} problem and rewrite it into the following optimization problem:
\begin{equation}\label{P1}
	\begin{split}
	\textbf{P1}: \ \min_{\boldsymbol{X}(t)} &\ \frac{1}{T}\sum_{t=1}^{T} \tau(t), \\
	&\text{s.t.} \quad \textbf{C1}\sim \textbf{C9},\\
	& \quad \textbf{C10'}:\ \lim\limits_{t \to \infty} \frac{\mathbb{E}{|Y_n(t)|}}{t}=0, \forall n \in \mathcal{N}.
	\end{split}
\end{equation}

To solve \textbf{P1}, we next transform the long-term stochastic problem \textbf{P1} into a static problem \textbf{P2} in each communication round by characterizing the Lyapunov drift-penalty function. 

\textbf{Definition 1 \cite{neely2010stochastic}}: Given $V>0$, the Lyapunov drift-penalty function is defined as 
\begin{equation}
\Delta_V(t) \triangleq \Delta \Omega(t)+V\tau(t),
\end{equation}
where $\Delta \Omega(t)\triangleq \mathbb{E}\{\Omega(t+1)-\Omega(t)|Y_n(t)\}$ is the conditional Lyapunov drift, and $\Omega(t) \triangleq \frac{1}{2}\sum_{n\in \mathcal{N}}{Y_n(t)}^2$ defines the Lyapunov function. 

Minimizing the drift-penalty function helps stabilize the virtual queue $Y_n(t)$, ensuring that it satisfies the average convergence rate stability constraint \textbf{C10'}. This, in turn, minimizes the delay in the FL system while also adhering to the segment scheduling and resource allocation constraints $\textbf{C1} \sim \textbf{C9}$ in a long-term perspective. The parameter $V$ is a control factor that balances the trade-off between delay minimization and the satisfaction of long-term convergence constraints. Therefore, we propose a dynamic segment scheduling and resource allocation algorithm to minimize the Lyapunov drift-penalty function. Using the conclusion of Lemma 1 in  \cite{xiumei2023LowLantencyFL_DNNPartition}, \textbf{P1} can be rewritten as
\begin{equation}\label{P2}
	\begin{split}
	\textbf{P2}: \ \min_{\boldsymbol{X}(t)} &\ V\tau(t)+\sum_{n\in \mathcal{N}}Y_n(t) \left(S(t)+p_n(t)\right),  \\
	&\text{s.t.} \quad \textbf{C1}\sim \textbf{C9}.
	\end{split}
\end{equation}

We summarize $\tau(t)$ as the following explicit expression:
\begin{equation}\label{tau_exp}
	\begin{split}
		\tau(t) &= (S+m-1)\times \\
		&\left( \frac{\delta_k (S o_l+mo'_l) }{m\phi_k f_k}+\frac{z^s+g^s}{B^{dd}\log_2 \left(1+\frac{p_k(t) h_{dd}(t)}{I_k+B_{dd}(t) N_0}\right)}  \right)\\
		&\quad +  \frac{\sum_{j=1}^{J}I_{n,j}(z_n^{enc}+\theta_n^{enc})}{B_n^U\log_2 \left(1+\frac{p_n(t) h_n(t)}{I_i+B_n^U(t) N_0}\right)}.
	\end{split}
\end{equation}

\subsection{Optimal Solution of P2}\label{s5.2}
To solve \textbf{P2}, we introduce two auxiliary variables, $\Lambda(t)$ and $\Upsilon(t)$. The first is to represent model partition and segment scheduling for each cluster performing TEB-wise pipeline parallel procedure:
\begin{equation}\label{auxiliary_lambada}
\begin{split}
	&\Lambda(t) =V \max_{k\in \mathcal{K}}\  \{\tau_{single}^{\text{pipe}}(t) \}+\sum_{n\in \mathcal{N}}Y_n(t)S(t) \\
	&=V(S+m-1) \max_{k\in \mathcal{K}}\left\{ \delta_k \frac{ S o_l+ mo'_l }{m\phi_k f_k}+r^{dd}_k(t) \right\}-Vr^{dd}_k(t)\\
	&\quad \quad \quad \quad +\sum_{n\in \mathcal{N}}Y_n(t)S(t).
\end{split}
\end{equation}

The variable $\Lambda_k(t)$ represents the total delay during D2D collaboration within one cluster if it is assigned by the model partition and segment scheduling strategy $\{\delta_k(t), S, m\}$ in the $t$-th communication round. 

In addition, we have $\Upsilon(t)$ of auxiliary variables for the FL procedure with D2E collaboration:
\begin{equation}\label{auxiliary_upsilon}
	\begin{split}
		\Upsilon(t) &=V\max_{n\in \mathcal{N}}\  \{\tau_n^{\text{Up}}(t)\}+\sum_{n\in \mathcal{N}}Y_n(t)p_n(t) \\
		&=V\max_{n\in \mathcal{N}} \left\{ \frac{z_n^{enc}+\theta_n^{enc}}{B_n^U(t)\log_2 \left(1+\frac{p_n(t) h_n(t)}{I_i+B_n^U(t) N_0}\right)} \right\} \\ 
		&\quad \quad \quad \quad+\sum_{n\in \mathcal{N}}Y_n(t)p_n(t).
	\end{split}
\end{equation}

As such, \textbf{P2} can be rewritten as 
\begin{equation}
	\begin{split}
	&\textbf{P3}: \min_{\boldsymbol{X}(t)} \ \Lambda(t)+\Upsilon(t), \\
	&\text{s.t.} \quad  \textbf{C1}\sim \textbf{C9}.
	\end{split}
\end{equation}

By exploiting the independence between $S(t), \Lambda(t)$ and $p_n(t),\tau_n^{\text{Up}}(t)$ in the objective function of \textbf{P3}, we decouple the joint optimization problem into the following sub-problems.

\begin{algorithm}[t!]
	\caption{Dynamic Segment Scheduling Algorithm}
	\begin{algorithmic}[1]
		\State \textbf{Input:} Control factor: $V$, Number of the TEBs: $L$.
		\State Initialize $S^{(0)}=1$, $m^{(0)}$, $\delta_k^{(0)}$, $\Gamma^{max}$;
		\State Optimize $\delta_k, S, m$ with block coordinate decent method;
		\Repeat
		\State Optimize $\delta_k$ and $S$ by solving (\ref{sub1-S_delta}) with integer programming;
		\State Optimize $m$ by solving (\ref{sub1-m});
		\State Update $Y_n(t)$ according to (\ref{queue});
		\Until{$Y_n(t)$ stabilizes}
		\State Compute the auxiliary variable $\Lambda(t)$ according to (\ref{auxiliary_lambada});
		\State \textbf{Return} $S^{\ast},m^{\ast},\delta_k^{\ast}$ and $\Lambda(t)$
	\end{algorithmic}
	\label{alg2}
\end{algorithm}

\subsubsection{Optimal Auxiliary Variable $\Lambda(t)$}
We minimize $\Lambda(t)$ by optimizing the segment number $S$, mini-batch size $m$, number of TEBs on each device $\delta_k$ for end-end collaboration. 

The first sub-problem formulation to minimize $\Lambda_k(t)$ is denoted as 
\begin{equation}\label{P3-lambda}
	\begin{split}
	 &\textbf{Sub-1}: \min_{\delta_k, S,m} \Lambda_k(t), \\
	&\text{s.t.} \quad  \textbf{C1}, \textbf{C2}, \textbf{C7}, \\
	& \quad \quad \textbf{C9'}: \frac{\delta_k(S o_l+mo'_l)}{m\phi_k}\left(f_k(t)\right)^2 +\frac{p_k(t) (z^s+g^s)}{r_k^{dd}(t)} \le E_{k}^{max}, \\
	& \quad \quad \quad \quad \forall n \in \mathcal{N}, \forall k \in \mathcal{K}.
	\end{split}
\end{equation}

Problem \textbf{Sub-1} involves integer and continuous variables, and both the objective function and constraints are non-convex. Thus, we use the alternating optimization (AO) method to reduce the difficulty of solving the problem. 
Since $S$ and $\delta_k$ are strongly coupled and both are integers, we determine group 1: $m$, group 2: $\delta$ and $S$. 
First we fix $S$ and $\delta_k$, and then optimize $m$, next the problem is transformed into:
\begin{equation}\label{sub1-m}
\begin{split}
&\min_{m} V(S+m-1)\max_{k\in \mathcal{K}} \left\{ \delta_k(t) \frac{S o_l + m o'_l}{m\phi_k f_k}+r^{dd}_k(t) \right\}, \\
&\text{s.t.} \quad \textbf{C7}, \textbf{C9'}, 
\end{split}
\end{equation}
and the optimal solution is obtained by taking the derivative of $m$.

Then we fix $m$, optimize $\delta_k$ and $S$. The problem is transformed into: 
\begin{equation}\label{sub1-S_delta}
\begin{split}
&\min_{\delta_k, S} V(S+m-1)\max_{k\in \mathcal{K}} \left\{ \delta_k(t) \frac{S o_l + m o'_l}{m\phi_k f_k}+r^{dd}_k(t) \right\} \\
& \quad \quad +S\sum_{n\in \mathcal{N}}Y_n(t), \\
&\text{s.t.} \quad \textbf{C1}, \textbf{C7}, \textbf{C9'}, \\
&  \quad  \quad \textbf{C2'}: z_k \le \delta_k \le M_{z_k}, S=\sum_{k} z_k, z_k=1_{\{\delta_k>0\}},\\
& \quad  \quad \textbf{C11}: \Gamma \le \Gamma_{max},
\end{split}
\end{equation}
where the \textbf{C2} is transformed into a linear constraint \textbf{C2'} through integer programming to obtain the optimal solution. 

The AO method allows us to optimize discrete variable $\delta_k$ and integer variables $S,m$ separately. The solution steps are summarized in Algorithm \ref{alg2}.

\begin{algorithm}[t!]
	\caption{Dynamic Resource Allocation Algorithm}
	\begin{algorithmic}[1]
		\State \textbf{Input:} Control factor $V$.
		\State Initialize $Y_n(t)=0$, $p^{(0)}_n$, $\Gamma^{max}$;
		\State \textbf{Require:} Channel state at the beginning of the $t$-th communication round;
		\State Optimize $I_{n,j},p_n$ with block coordinate decent method;
		\Repeat
		\State Optimize $I_{n,j}(t)$ by solving (\ref{sub2-I}) with improved Hungarian method;
		\State Linearize Constraint \textbf{C8'} and \textbf{C11} and optimize $p_n(t)$ by solving (\ref{sub2-p_n}) with SCA method;
		\State Update $Y_n(t)$ according to (\ref{queue});
		\Until{$Y_n(t)$ stabilizes}
		\State Compute the auxiliary variable $\Upsilon(t)$ according to (\ref{auxiliary_upsilon}); 
		\State \textbf{Return} $I^{\ast}(t),p^{\ast}_n(t)$ and $\Upsilon(t)$
	\end{algorithmic}
	\label{alg3}
\end{algorithm}

\subsubsection{Optimal Auxiliary Variable $\Upsilon(t)$}
The second sub-problem formulation for D2E communication is denoted as 
\begin{equation}\label{sub2}
\begin{split}
		&\textbf{Sub-2}: \min_{I_{n,j},p_n}\ V\frac{\sum_{j=1}^{J}I_{n,j}(t)(z_n^{enc}+\theta_n^{enc})}{B_n^U\log_2 \left(1+\frac{p_n(t) h_n}{I_i+B_n^U N_0}\right)} \\ 
		&\quad \quad \quad \quad \quad \quad \quad \quad \quad \quad \quad \quad \quad +\sum_{n\in \mathcal{N}}Y_n(t)p_n(t), \\
		\text{s.t.} &\quad \textbf{C3}\sim \textbf{C6},\\
		&\quad \textbf{C8'}: \frac{p_n  \theta_n^{enc}}{B_n^U \left(\log_2 \left(1+\frac{p_n h_n}{I_i+B_n^U N_0}\right)\right)}  \le E_n^{max}, \forall n \in \mathcal{N}.
\end{split}
\end{equation}

We also use the AO method to solve Problem \textbf{Sub-2}, where the group 1 is $p_n$, the group 2 is $I_{n,j}$. 

Firstly, we solve the channel matching problem. Fixing $p_n(t)$, Problem Sub-2 is transformed as a Weighted matching problem:
\begin{equation}\label{sub2-I}
\min_{I_{n,j}} \sum_{j\in \mathcal{J},n\in\mathcal{N}}I_{n,j}\left[\frac{V(z_n^{enc}+\theta^{enc}_n)}{B^{U}_n\log_2(\text{SINR}_n)+Y_n(t)p_n}\right].
\end{equation}

In Problem (\ref{sub2-I}), some CUs cannot be allocated channels due to resource limitations, so we use the improved Hungarian algorithm to construct a virtual channel ($I_{n,j}=0$) to solve the bipartite graph matching problem.

Next, we solve the power control problem. Fixing $I_{n,j}$, the objective involves fractions and logarithmic functions. We adopt Dinkelbach Transformation and Successive Convex Approximation (SCA) method \cite{razaviyayn2014SuccessiveConvexApproxima}, then Problem \textbf{Sub-2} is transformed into:
\begin{equation}\label{sub2-p_n}
\begin{split}
&\min_{p_n} V\sum_{j\in \mathcal{J}}I_{n,j}(z_n^{enc}+\theta_n^{enc}) \\
&\quad \quad -\lambda B^U\log_2\left(1+\frac{p_nh_n}{I_i+B^U_nN_0}\right)+\sum_{n\in \mathcal{N}}Y_np_n \\
\text{s.t.} &\quad \textbf{C3}\sim \textbf{C6}, \textbf{C8'}, \\
&\quad \textbf{C11}: \Gamma \le \Gamma_{max}.
	\end{split}
\end{equation}

We adopt the first-order Taylor approximation for convex approximation of \textbf{C8'}, and the gradient of the function with respect to $p_n$ is calculated to construct a convex approximate constraint. At the $i$-th iteration, given the current power $p_n$, the approximate constraint is
\begin{equation}
\textbf{C8''}: p_n \le \frac{E^{max}_nB^{U}_n[f(p_n)-p_n \nabla f(p_n)]}{\theta^{enc}_n-E^{max}_nB^{U}_n\nabla f(p_n)},
\end{equation}
where 
\begin{equation*}
\begin{split}
&f(p_n)=\mathbb{E}_{h_n}\left[\log_2\left(1+\frac{p_nh_n}{I_i+B^{U}_nN+0}\right)\right], \\
&\nabla f(p_n)=\mathbb{E}_{h_n}\left[\frac{h_n}{(I_i+B^U_nN_0+p_nh_n)\ln2}\right].
\end{split}
\end{equation*}

Similarly, we linearize constraint \textbf{C11}, Taylor expansion $\frac{C}{p_nh_n+I_i}$ at point $p_n^{(e)}$, then the linear approximation constraint of $\Gamma$ is:
\begin{equation}
\textbf{C11'}: \frac{\beta \eta^2}{2N} \left[\frac{\phi^2 }{L}S^2 +[g_1(p_n)+g_2(p_n)]+\phi^2 \right] \le \Gamma^{max},
\end{equation}
where $g_1$ and $g_2$ are the first and second terms of $\Gamma$, respectively, after Taylor expansion at $p_n^{(e)}$, i.e.,
\begin{equation*}
g_1(p_n)=\frac{C}{p^{(e)}_nh_n+I_i},
\end{equation*}
\begin{equation*}
g_2(p_n)=-\frac{Ch_n}{(p_n^{(e)}h_n+I_i)^2}(p_n-p^{(e)}_n).
\end{equation*}

As such, we use the convex optimization tool such as CVXPY to solve the problem (\ref{sub2-p_n}).

According to the current solution $p^{\ast}$ and $S^{\ast}$, we calculate the value of $\Gamma^t$ and update the queue $Y_n(t)$ in Eq. (\ref{queue}). The solution steps are summarized in Algorithm \ref{alg3}.

Given the optimized $\Lambda(t)$ and $\Upsilon(t)$, we obtain the optimal scheduling protocol $\textbf{X}^{\ast}(t)$. Algorithm \ref{alg4} shows the Dynamic online algorithm for Segment Scheduling and Resource Allocation (DSSRA).

\begin{algorithm}[t!]
	\caption{Dynamic Segment Scheduling and Resource Allocation (DSSRA)}
	\begin{algorithmic}[1]
		\State \textbf{Input:} Control factor $V$.
		\State Initialize $t=0$;
		\For{each communication $t=1,2,...,T$}
		\Repeat
		\State Compute $\Lambda(t)$ and $\Upsilon(t)$ according to (\ref{auxiliary_lambada}) and (\ref{auxiliary_upsilon}) with the current optimal $\textbf{X}^{\ast}(t)$;
		\State Given the optimized auxiliary variables $\Lambda(t)$ and $\Upsilon(t)$, find the optimal segment number $S$ and power $p_n$ with block coordinate descent methd;
		\State Update $Y_n(t)$ according to (\ref{queue});
		\Until{$Y_n(t)$ stabilizes}
		\EndFor
		\State \textbf{Return} $\textbf{X}^{\ast}(t)$
	\end{algorithmic}
	\label{alg4}
\end{algorithm}

\subsection{Complexity Analysis}\label{s5.3}
We analyze the complexity of the proposed DSSRA algorithm. In Algorithm 2, optimizing $m$ requires a closed-form solution with a complexity of $\mathcal{O}(K)$. Optimizing $S$ requires integer programming with constraints, leading to a worst-case complexity of $\mathcal{O}(2^K)$; however, this is acceptable when $K \le 10$. Therefore, the alternating AO solution has an overall per-iteration complexity of approximately $\mathcal{O}(K^3)$.

In Algorithm 3, the Hungarian algorithm for channel matching has a complexity of $\mathcal{O}(J^3)$, while the SCA method for power allocation requires gradient calculations in each iteration, with a complexity of $\mathcal{O}(N)$.

Algorithm 4 calls Algorithms 2 and 3 once per round and updates the virtual queue. Since the virtual queue update has a complexity of $\mathcal{O}(N)$, the total time complexity per round is $\mathcal{O}(K^3 + J^3 + N)$. The overall space complexity, which includes device allocation, the channel allocation matrix, the virtual queue, temporary gradients, and latency, is $\mathcal{O}(NJ + K)$.

\section{Experiments}\label{s6}
The experiments were conducted on different tasks: machine translation, named entity recognition and sentence classification. To ensure a fair comparison, we keep the data distribution and training settings as consistent as possible with other learning framework.

\subsection{Experimental Setup}\label{s6.1}
\subsubsection{Models}
To assess the performance of CollaPipe for the Transformer-based LLM training, we conduct experiments using both the classic Transformer model and the BERT model, which are widely applied in natural language understanding tasks.
\begin{itemize}
	\item Classic Transformer \cite{vaswani2017attention}: The classic Transformer model comprises two main components: the Encoder and the Decoder, each consisting of 6 Transformer blocks. The decoder and encoder have similar structures, but their sublayers are different, and the interactions between decoder blocks are more complex.
	\item BERT \cite{devlin2019bert}: BERT consists of 12 Transformer encoder blocks and is a large language model with an encoder-only architecture.
\end{itemize}

\subsubsection{Benchmark}
To evaluate the performance of CollaPipe, we conduct three experiments: (i) training the classic Transformer model on the Multi30K dataset for a machine translation task, (ii) pre-training the BERT model on a Chinese named entity recognition (NER) dataset, and (iii) fine-tuning the BERT model on a 15-class Chinese dataset to assess accuracy in a sentence classification task.\footnote{All datasets are available at: https://github.com/moon-hotel/BertWithPretrained.}

For comparison, we adopt three baseline methods:
\begin{itemize}
\item VanillaFL: A standard FL approach that uses simple averaging for model aggregation, without any model partitioning or parallel optimization.

\item PipeLine: A batch-based pipeline parallelism method that does not perform model segmentation optimization or apply FL aggregation.

\item TITANIC \cite{su2024titanic}: A model auto-splitting framework that incorporates device selection strategies and uses a P2P communication protocol.
\end{itemize}

To further evaluate the effectiveness of our dynamic online scheduling algorithm, we compare it with the following additional strategies:
\begin{itemize}
	\item Random Scheduling: In each round of communication, BS randomly selects $J$ CUs for communication, and each CU randomly determines the training order of its associated devices.
	\item Loss-only Driven Scheduling: The BS selects $J$ CUs based on the training loss of local model in each cluster.
	\item Delay-only Driven Scheduling: The BS selects $J$ CUs based on the local training delay reported by each cluster.
\end{itemize}

\subsubsection{Configurations}
We consider a default setting with $J=4$ channels, $N=3$ clusters, each comprising $K=6$ devices. To estimate the theoretical computation and scheduling time for each segment, we use a predefined formula to calculate FLOPs. To mitigate hardware-level variations in computation time, we leverage the \texttt{Transformers} and \texttt{fvcore} libraries to precompute the FLOPs for each segment.
More settings for CollaPipe and DSSRA algorithms are shown in Table \ref{etab2}.

\begin{table}[tb]
	\caption{Experimental parameter setting for our scheme \label{etab2}}
	\centering
	\begin{tabular}{|>{\centering\arraybackslash}m{1cm}|>{\centering\arraybackslash}m{2.2cm}| |>{\centering\arraybackslash}m{1cm}|>{\centering\arraybackslash}m{2.2cm}|}
		\hline
		{\bf{Symbols}} & {\bf{Value}} & {\bf{Symbols}} & {\bf{Value}}\\
		\hline
		$E^{max}_n$ & 10 J & $E^{max}_k$ & 5 J \\
		\hline
		$\gamma^{max}_n$ & 3 G & $B^U_n$ & [0.4, 0.6] MHz \\
		\hline
		$\gamma^{max}_k$ & 1.5 G & $B^{dd}$ & 0.5 MHz  \\
		\hline
		$f_k$ & [0.1, 0.8] GHz &  $p_k$ & [0.07, 0.1] W \\
		\hline
		$P^{max}_n$ & 0.5 W & $N_0$ & -174 dBm/Hz \\
		\hline
		$P^{max}_k$ & 0.18 W & $V$ & [0.01, 10, 100]   \\
		\hline
		$h_{dd}$ & -30 & $I_{dd}$ & $5\times 10^{-10}$ W \\
		\hline
		$h_n$ & $[-0.12, -0.08]$ & $I_i$ & $[0.06, 0.08]$ W  \\
		\hline
		$b$ & 64 &  $\phi_k$ & [10, 24] MFLOPs  \\
		\hline
		$o_l$ & $2 \times 10^6$ FLOPs & $o'_l$ & $2 \times 10^6$ FLOPs \\
		\hline
	\end{tabular}
\end{table}

\subsection{Evaluation within one cluster}
\begin{figure}[t]
	\centering{\includegraphics[width=0.35\textwidth]{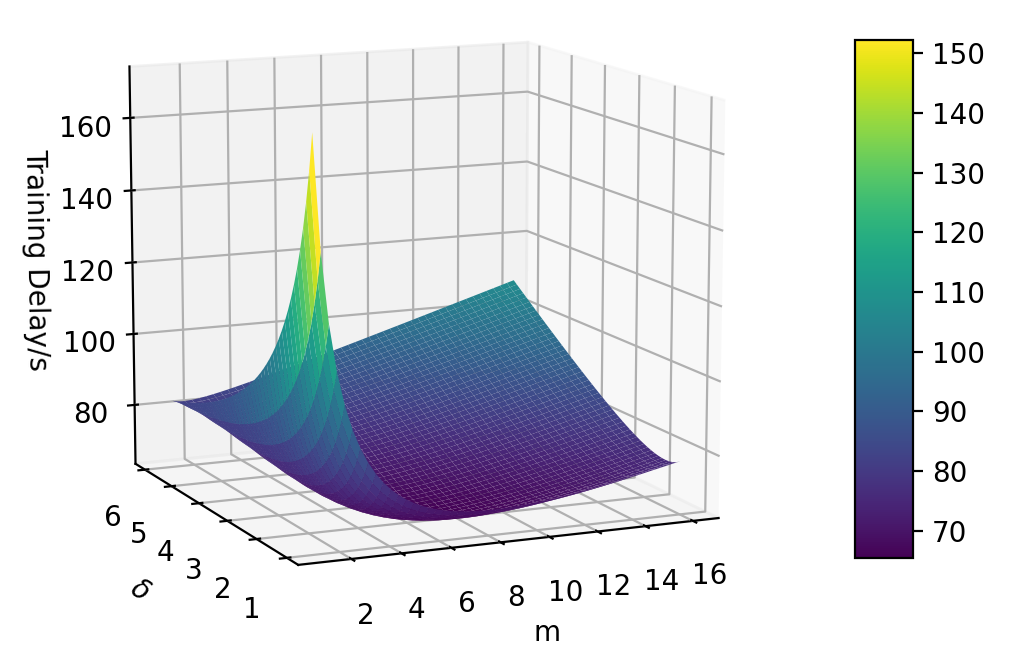}}
	\caption{Training time with the even distribution of $\delta$.}
	\label{fig4}
\end{figure}

To verify the impact of the number of segments $S$ and micro-batches $m$, we first test the device latency under uniform partitioning, i.e., $S=\frac{L}{\delta_k}$. Taking Transformer model as an example, we evaluate different configurations. As can be seen from Fig. \ref{fig4}, if the model is evenly split and $S$ devices are selected to execute the pipeline parallel strategy, there is always an optimal $S$ and $m$ value. However, in the mobile edge scenario, the resource conditions of the devices are different.

\begin{figure}[!t]
	\centering{\includegraphics[width=0.3\textwidth]{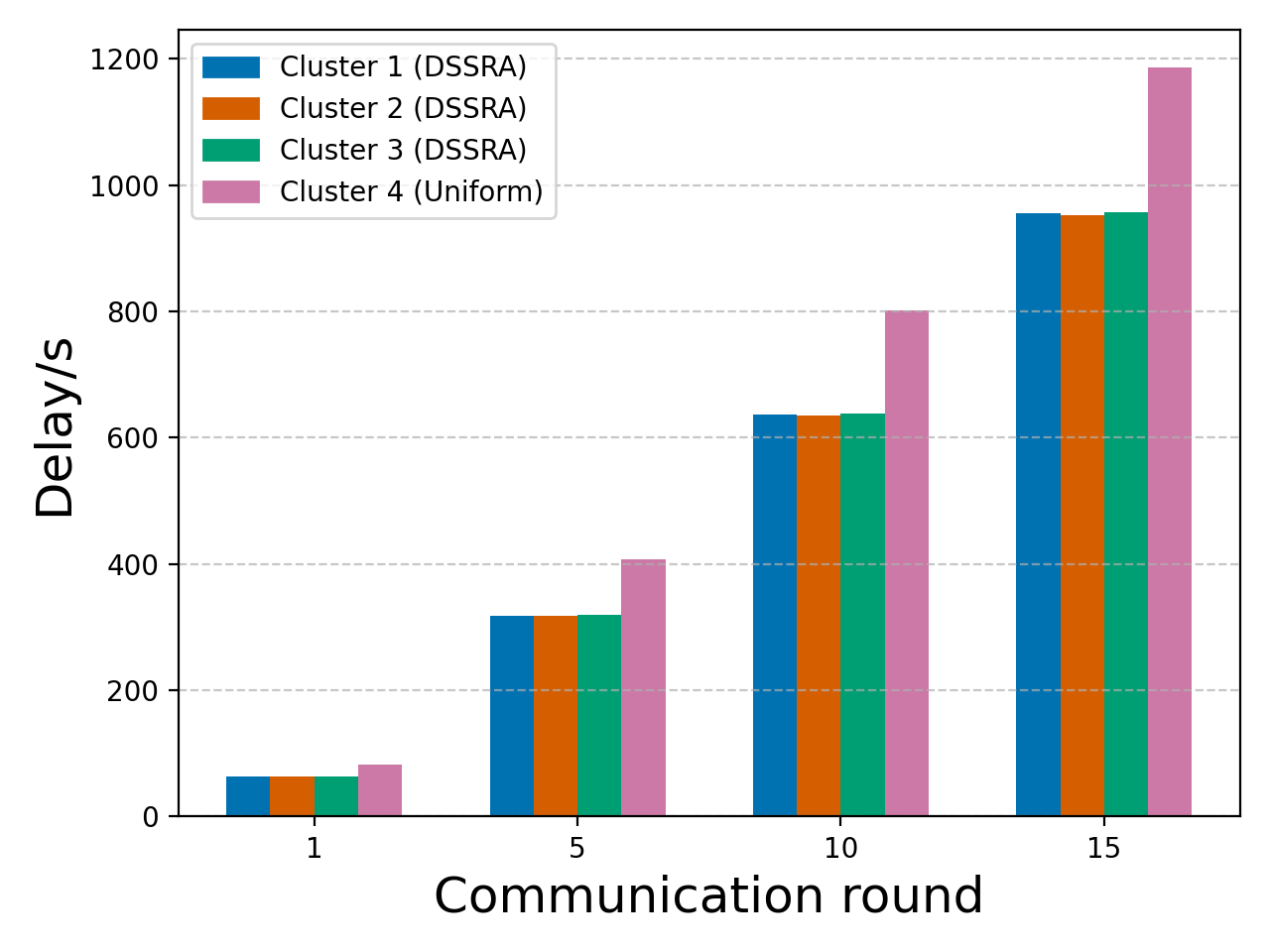}}
	\caption{Training delay of one cluster with heterogeneous devices.}
	\label{fig5}
\end{figure}

We simulate a heterogeneous edge network environment to evaluate the training delay of the BERT model. As shown in Fig.~\ref{fig5}, the DSSRA algorithm dynamically determines the optimal number of model segments and their corresponding assignments to mobile devices, effectively balancing computation and communication delays across devices with varying capabilities.
For clusters with imbalanced computational resources, the optimal scheduling strategy yields the configuration $S=\{5,4,3\}, m=\{14,11,9\}$. Fig.~\ref{fig5} also presents a comparative scenario in which one cluster (cluster 4) deviates from the proposed scheduling strategy and instead adopts a uniform TEB allocation policy. In this case, training latency increases significantly, highlighting the effectiveness of the proposed adaptive scheduling mechanism. These results provide empirical support for the validity of Assumption 4 of our system model.

\begin{figure*}[!t]
	\centering
	\begin{subfigure}[t]{0.3\textwidth}
		\centering
		\includegraphics[width=\textwidth]{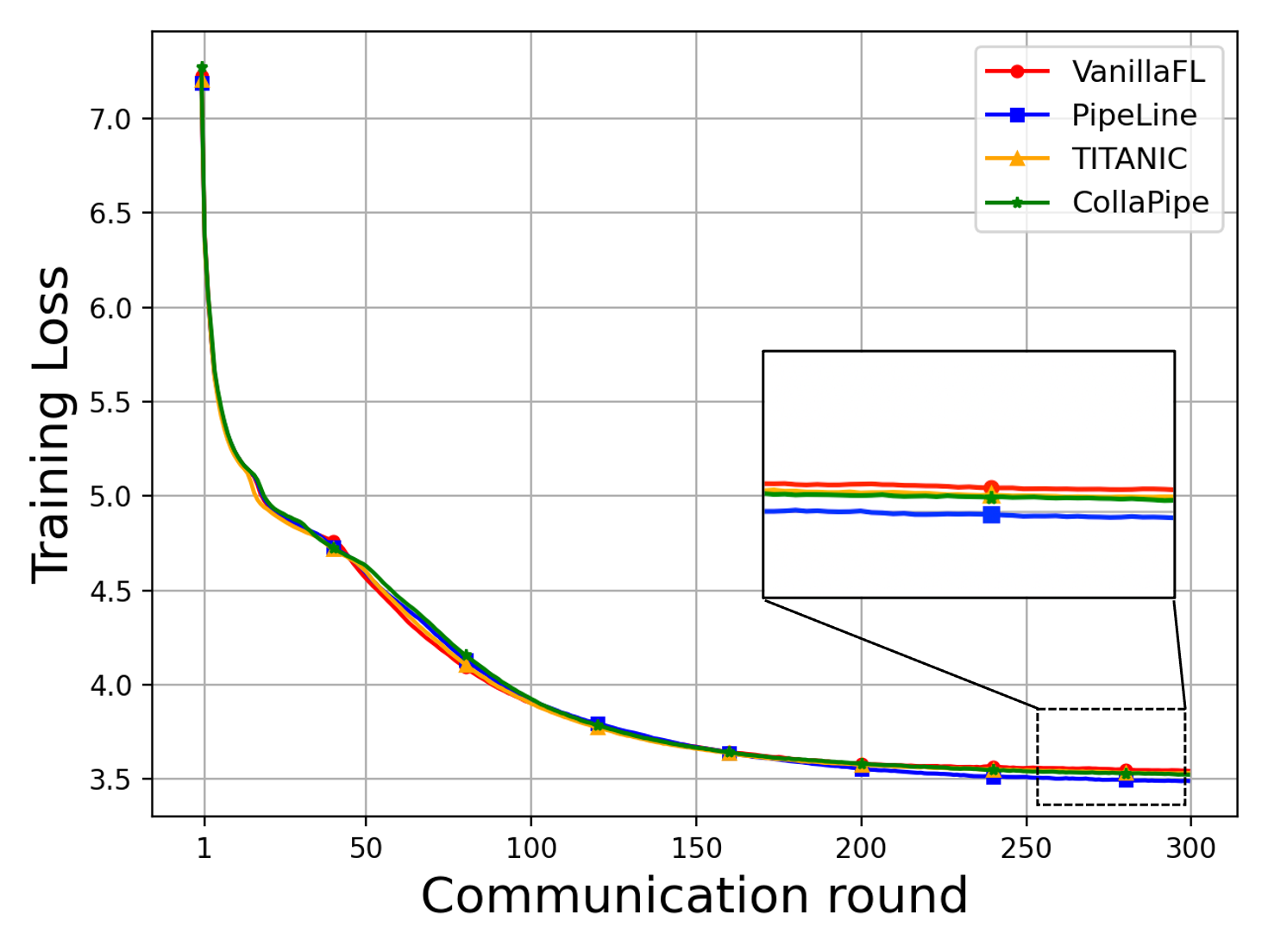}
		\caption{Training loss of Machine Translation}
		\label{fig6-a}
	\end{subfigure}
	\hspace{1em}  % ← 调整子图间距
	\begin{subfigure}[t]{0.3\textwidth}
		\centering
		\includegraphics[width=\textwidth]{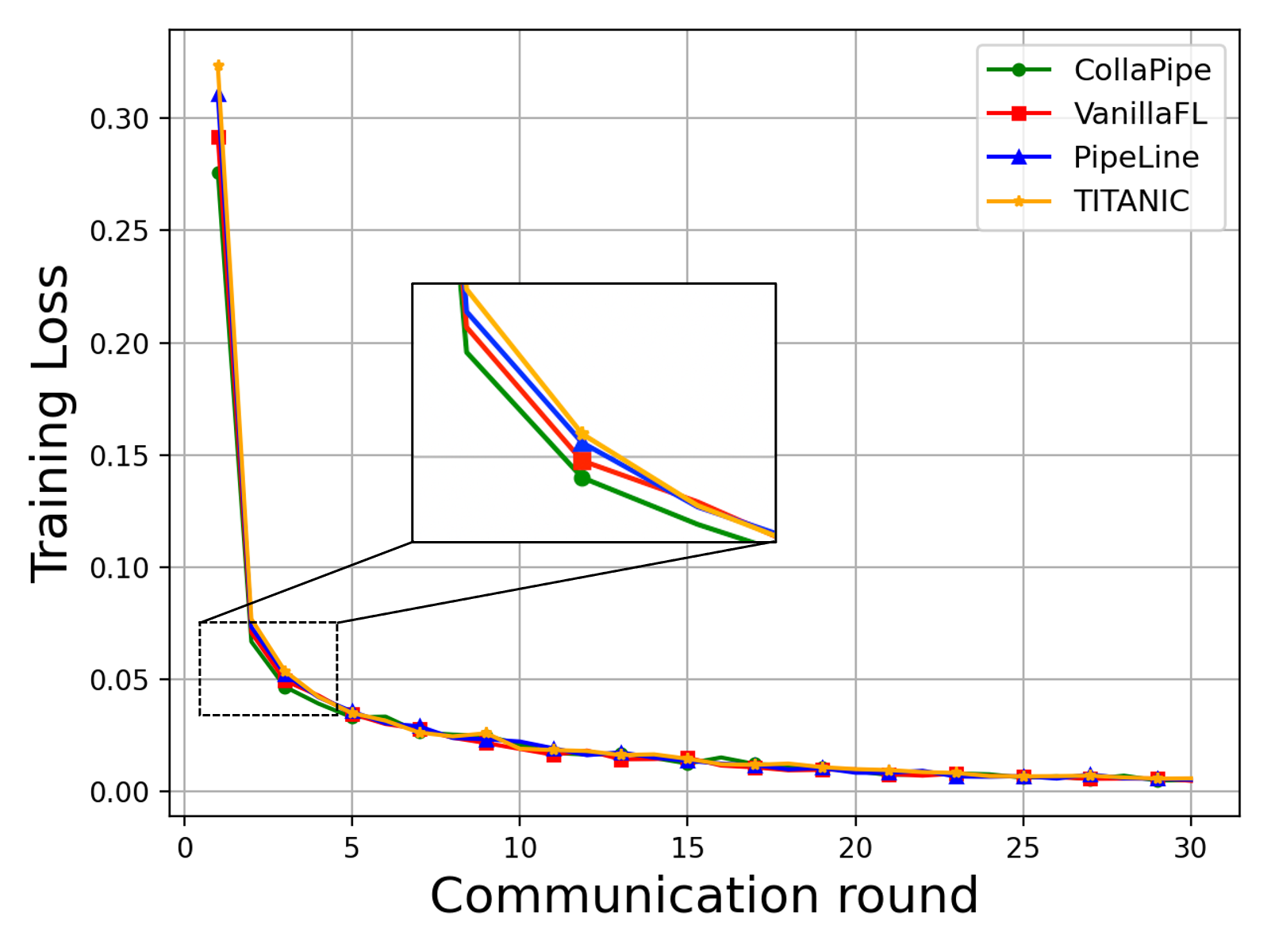}
		\caption{Training loss of Named Entity Recognition}
		\label{fig6-b}
	\end{subfigure}
	\hspace{1em}  % ← 调整子图间距
	\begin{subfigure}[t]{0.3\textwidth}
		\centering
		\includegraphics[width=\textwidth]{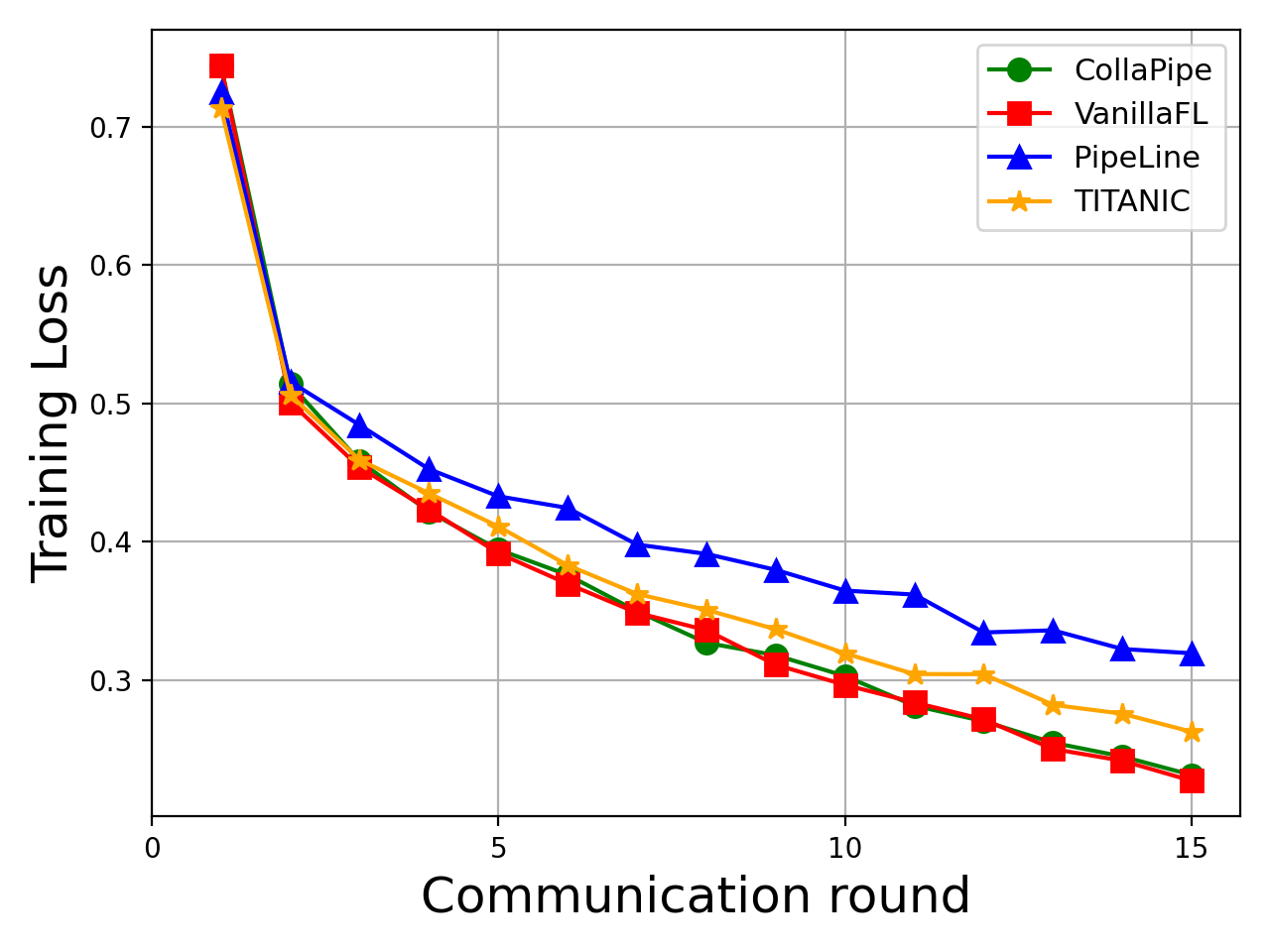}
		\caption{Training loss of Sentence Classification}
		\label{fig6-c}
	\end{subfigure}
	\hspace{1em}  % ← 调整子图间距
	\begin{subfigure}[t]{0.3\textwidth}
		\centering
		\includegraphics[width=\textwidth]{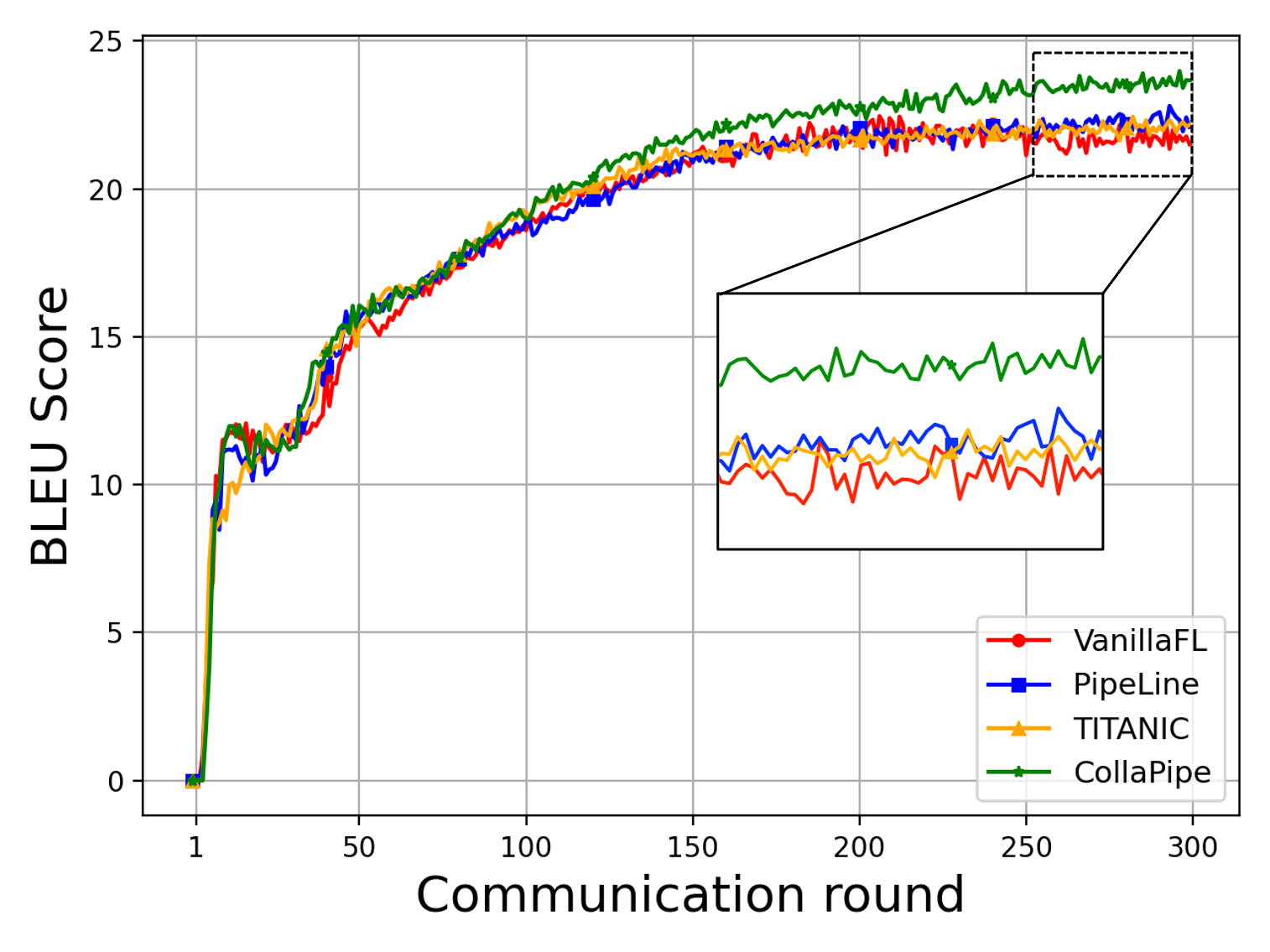}
		\caption{BLEU score of Machine Translation}
		\label{fig6-d}
	\end{subfigure}
	\hspace{1em}  % ← 调整子图间距
	\begin{subfigure}[t]{0.3\textwidth}
		\centering
		\includegraphics[width=\textwidth]{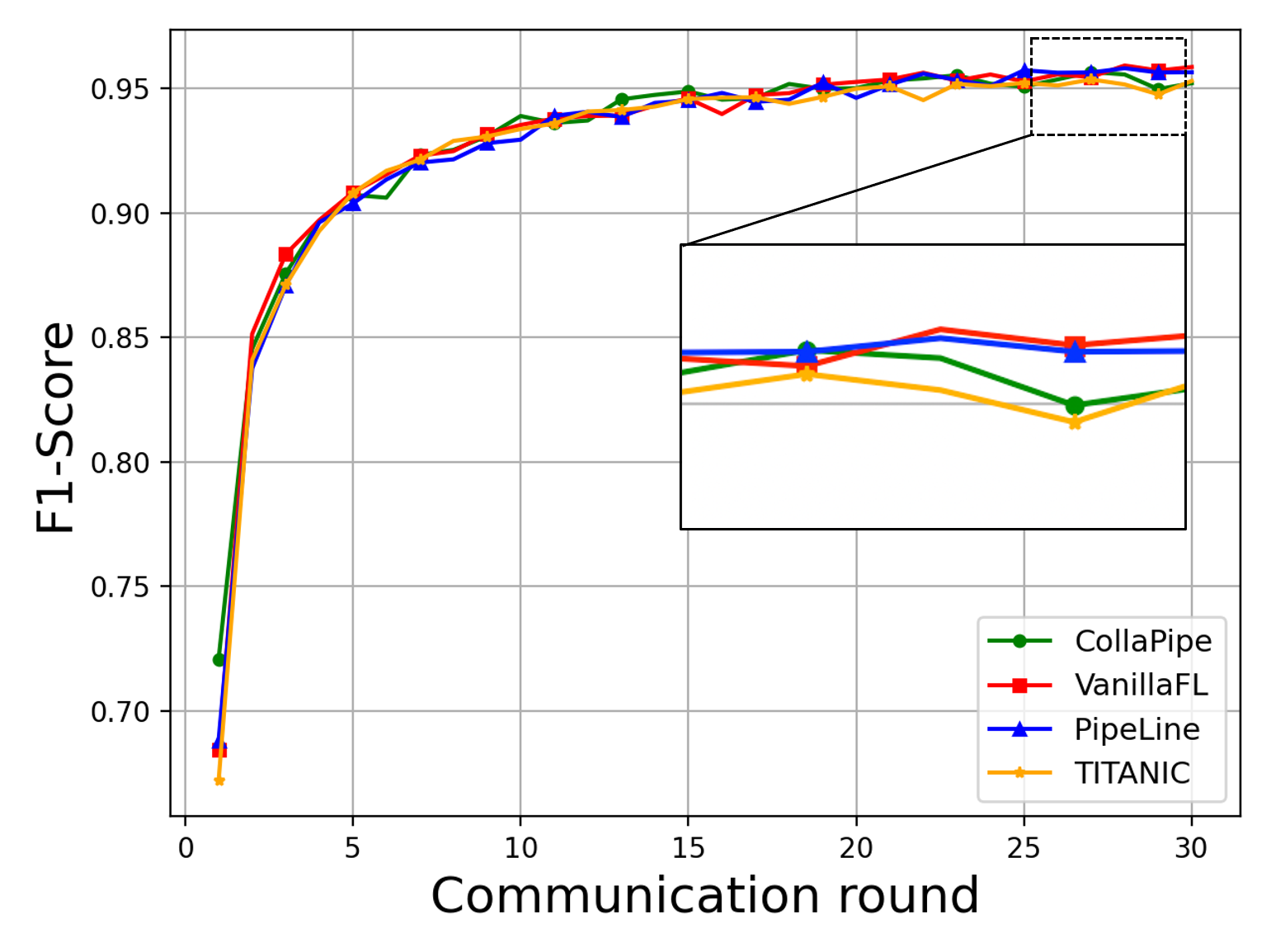}
		\caption{F1 score of Named Entity Recognition}
		\label{fig6-e}
	\end{subfigure}
	\hspace{1em}  % ← 调整子图间距
	\begin{subfigure}[t]{0.3\textwidth}
		\centering
		\includegraphics[width=\textwidth]{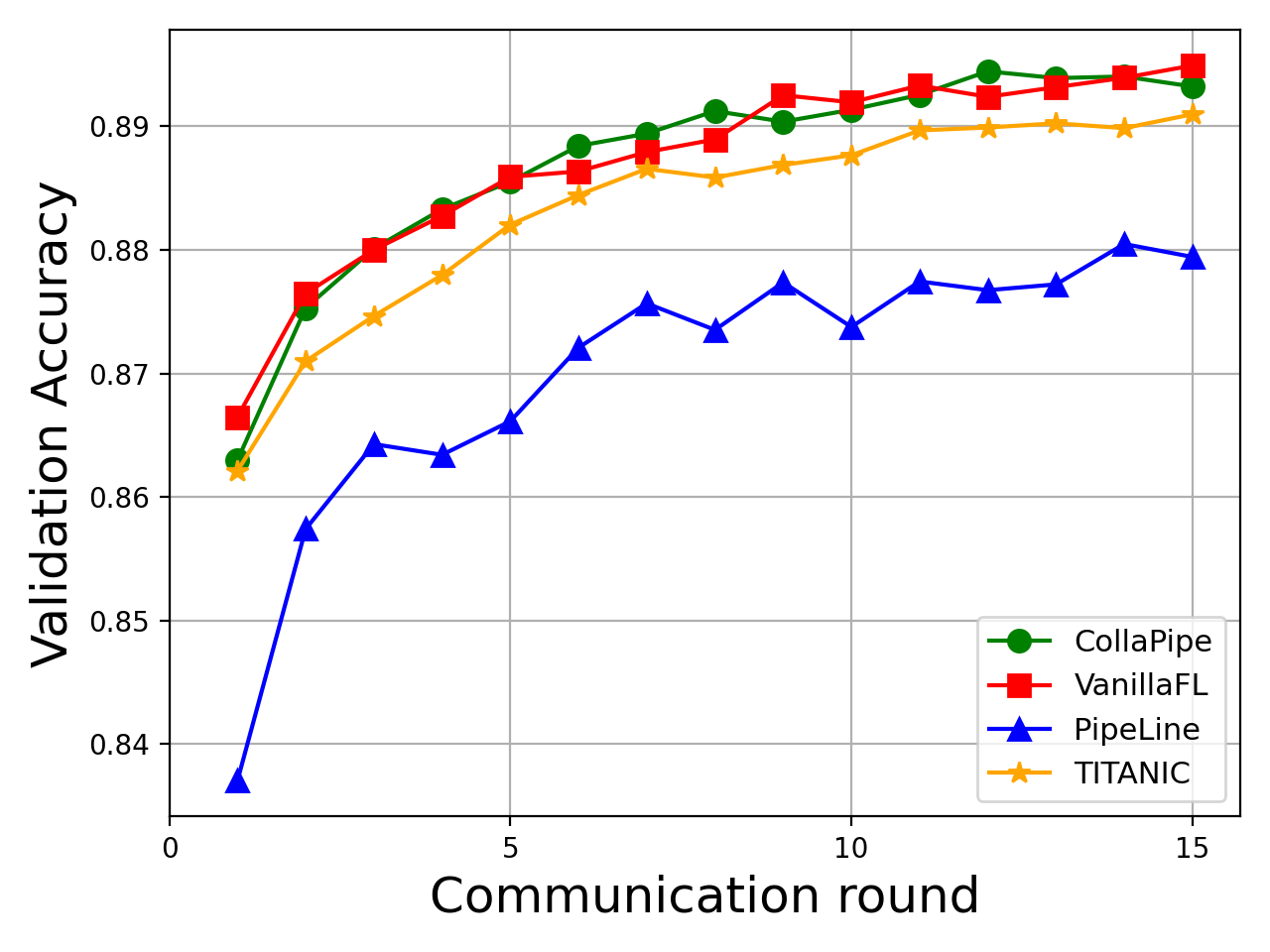}
		\caption{Accuracy of Sentence Classification}
		\label{fig6-f}
	\end{subfigure}
	\caption{Learning performance of various tasks with the Transformer and BERT.}
	\label{fig6}
\end{figure*}

\subsection{Evaluation of the CollaPipe Framework}
We first pre-trained the original Transformer model under our proposed scheme and compared it with the baselines. To ensure fairness in the comparison, all distributed learning frameworks use the same configuration.

The learning performance under different methods is illustrated in Fig.~\ref{fig6}. As shown in Fig. \ref{fig6-a}, under identical batch size and channel conditions, no significant differences are observed in the training loss across the compared methods. However, we can see from Fig. \ref{fig6-d} that our optimized framework can better adapt to machine translation tasks, with the BLEU score improved by 19.5\% compared to baselines.

The BERT model is pre-trained on the named entity recognition (NER) task, and the corresponding learning performance is illustrated in Figs.~\ref{fig6-b} and \ref{fig6-e}. The proposed framework achieves a faster reduction in training loss during the early stages of training, indicating improved convergence speed and learning efficiency in the initial phase.

Based on the pre-trained model, we fine-tune the BERT model for a single-sentence classification task. 
The training process converges more rapidly—typically within 15 training rounds, as shown in Fig.~\ref{fig6-c}. Compared with the existing resource scheduling strategies in federated learning systems, the proposed method does not show a significant advantage in learning performance. However, it consistently outperforms the PipeLine and TITANIC mechanisms by about 1.82\% and 0.15\%, respectively.
These results demonstrate that the proposed learning framework, under independently and identically distributed (IID) data conditions, effectively compensates for the training loss caused by encoder partitioning through adaptive device scheduling and resource management. As shown in Figs.~\ref{fig6-a}, \ref{fig6-b}, and \ref{fig6-c}, this approach maintains a relatively stable training loss. 
Moreover, we observe that CollaPipe achieves more significant performance gains in machine translation and sentence classification tasks.
We conjecture that these improvements may stem from differences in block-level or token-level behavior introduced by model partitioning. 

\begin{figure}[t!]
	\centering
	\includegraphics[width=2.5 in]{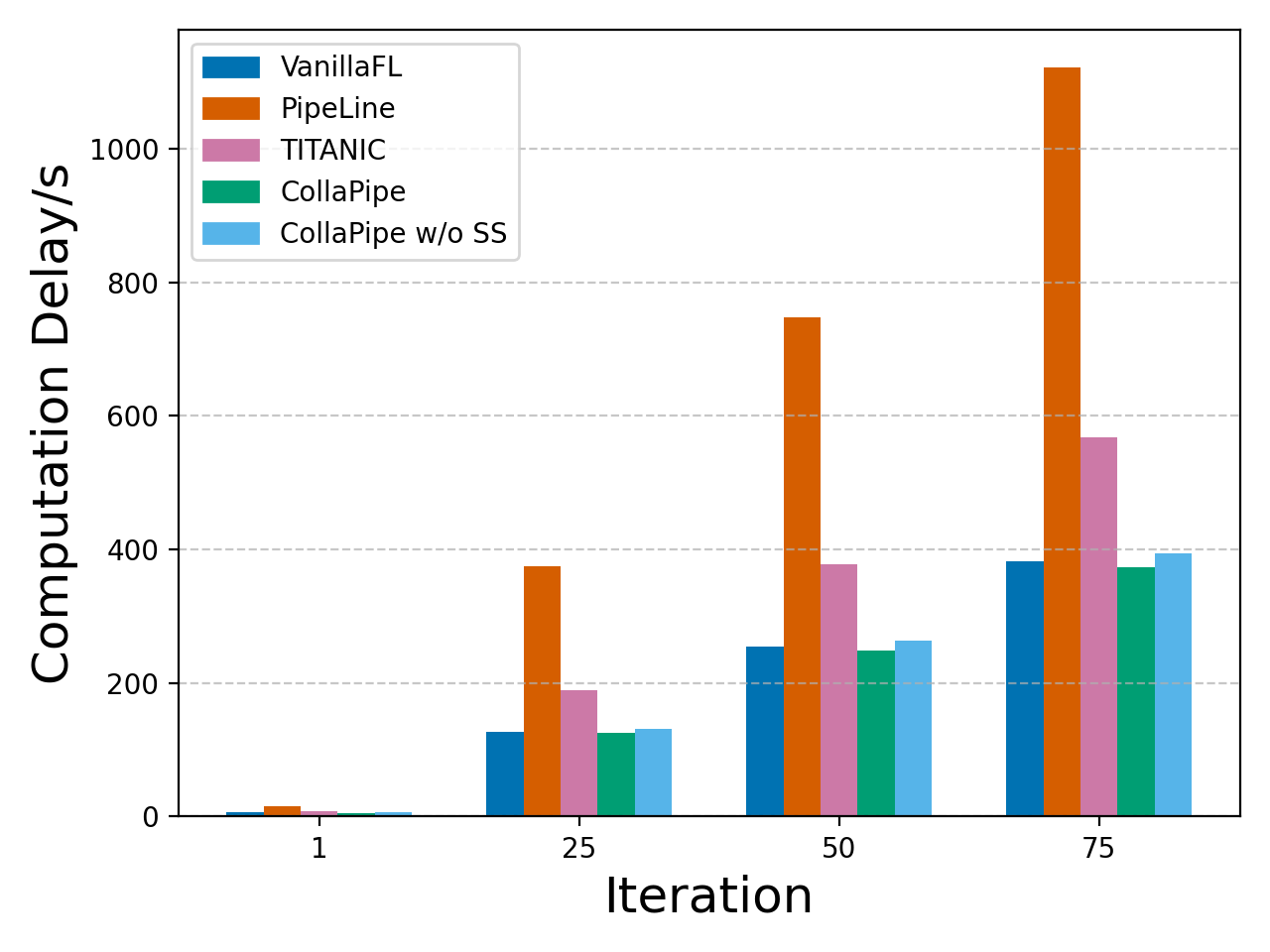}
	\caption{On-device accumulated computation time of machine translation with the Transformer.}
	\label{fig7}
\end{figure}

\begin{figure}[t!]
	\centering
	\includegraphics[width=2.5 in]{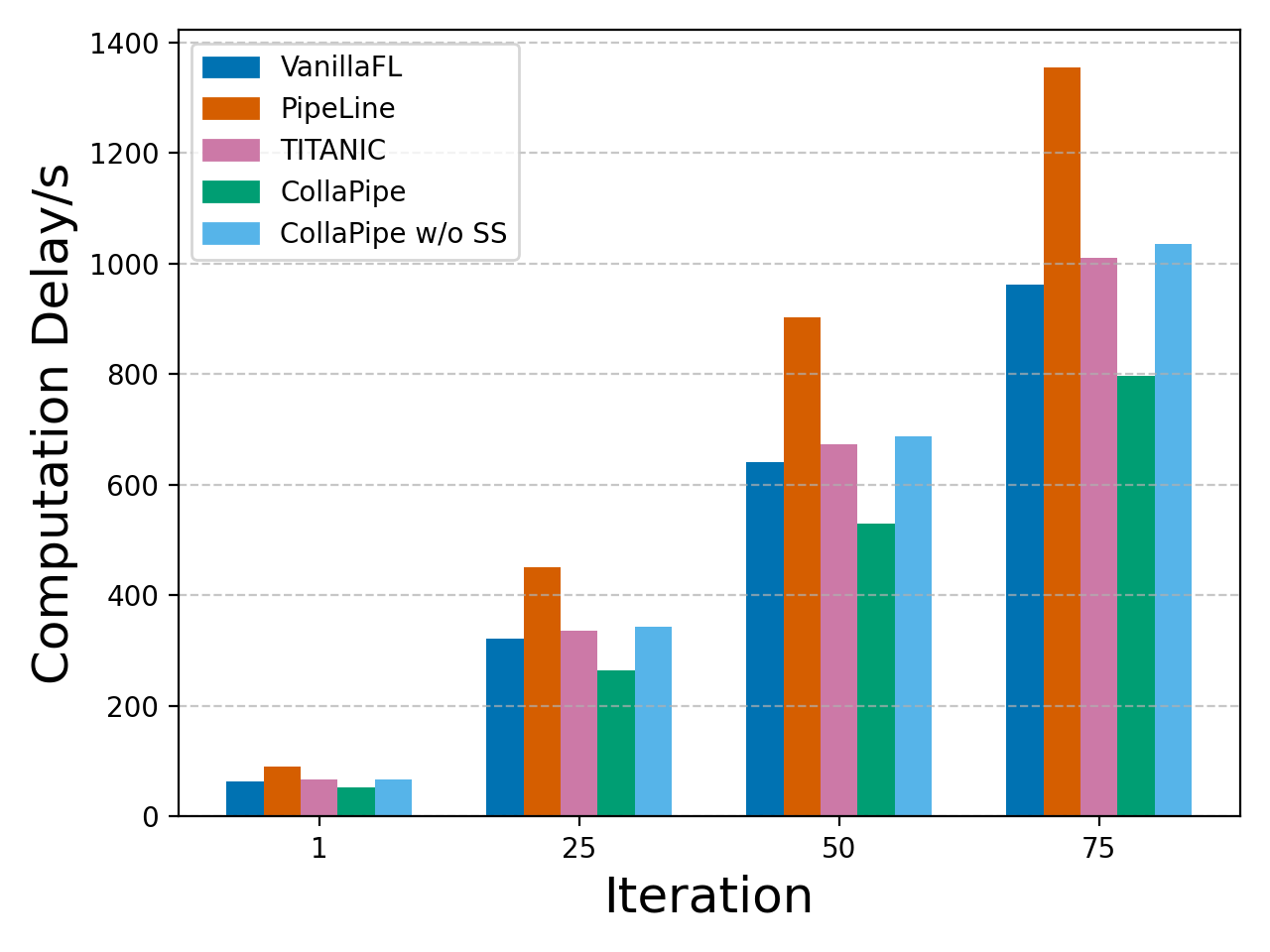}
	\caption{On-device accumulated computation time of sentence classification with the BERT.}
	\label{fig8}
\end{figure}

\begin{table}[t!]
	\caption{Scalability of different training frameworks (i.e., the BERT model) \label{etab3}}
	\centering
	\begin{tabular}{|>{\centering\arraybackslash}m{1.5cm}|>{\centering\arraybackslash}m{1.5cm}|>{\centering\arraybackslash}m{1.5cm}|>{\centering\arraybackslash}m{1cm}|>{\centering\arraybackslash}m{1cm}|}
		\hline
		Method & Memory usage of single devices & Number of data users & Segment Scheduling & Resource Allocation \\
		\hline
		Vanilla FL & $\sim$5-6 G & Large & N & N \\
		\hline
		PipeLine & $\sim$5-6 G  & Small & N & N \\
		\hline
		TITANIC & Adapted  & Large & N & Y \\
		\hline
		\textbf{CollaPipe} & \textbf{Adapted} & \textbf{Small} & \textbf{Y} & \textbf{Y} \\
		\hline
	\end{tabular}
\end{table}

Fig. \ref{fig7} presents the accumulated computation time during Transformer training. CollaPipe consistently achieves the lowest computational latency across all benchmark methods.
Fig. \ref{fig8} shows that VanillaFL achieves a 29.98\% reduction in training delay compared to the PipeLine method. In contrast, other approaches introduce additional scheduling latency due to the need for inter-device coordination.
Among all evaluated schemes, CollaPipe achieves the best performance, reducing training delay by 18.94\% compared to TITANIC, 15.09\% compared to VanillaFL, and 40.55\% compared to PipeLine. However, when the segment scheduling module is removed from CollaPipe, latency increases by approximately 22.45\%, indicating a significant drop in efficiency.
The light blue bars in the figures represent CollaPipe without the segment scheduling module (CollaPipe w/o SS). Compared to the fully optimized framework, the cumulative computation latency increases significantly, demonstrating that the proposed DSSRA algorithm effectively reduces intra-cluster computation delay.
These results highlight the effectiveness of CollaPipe's coordinated scheduling mechanism in minimizing system delay and improving training efficiency in distributed learning environments.

\begin{figure}[t!]
	\centering
	\includegraphics[width=2.5 in]{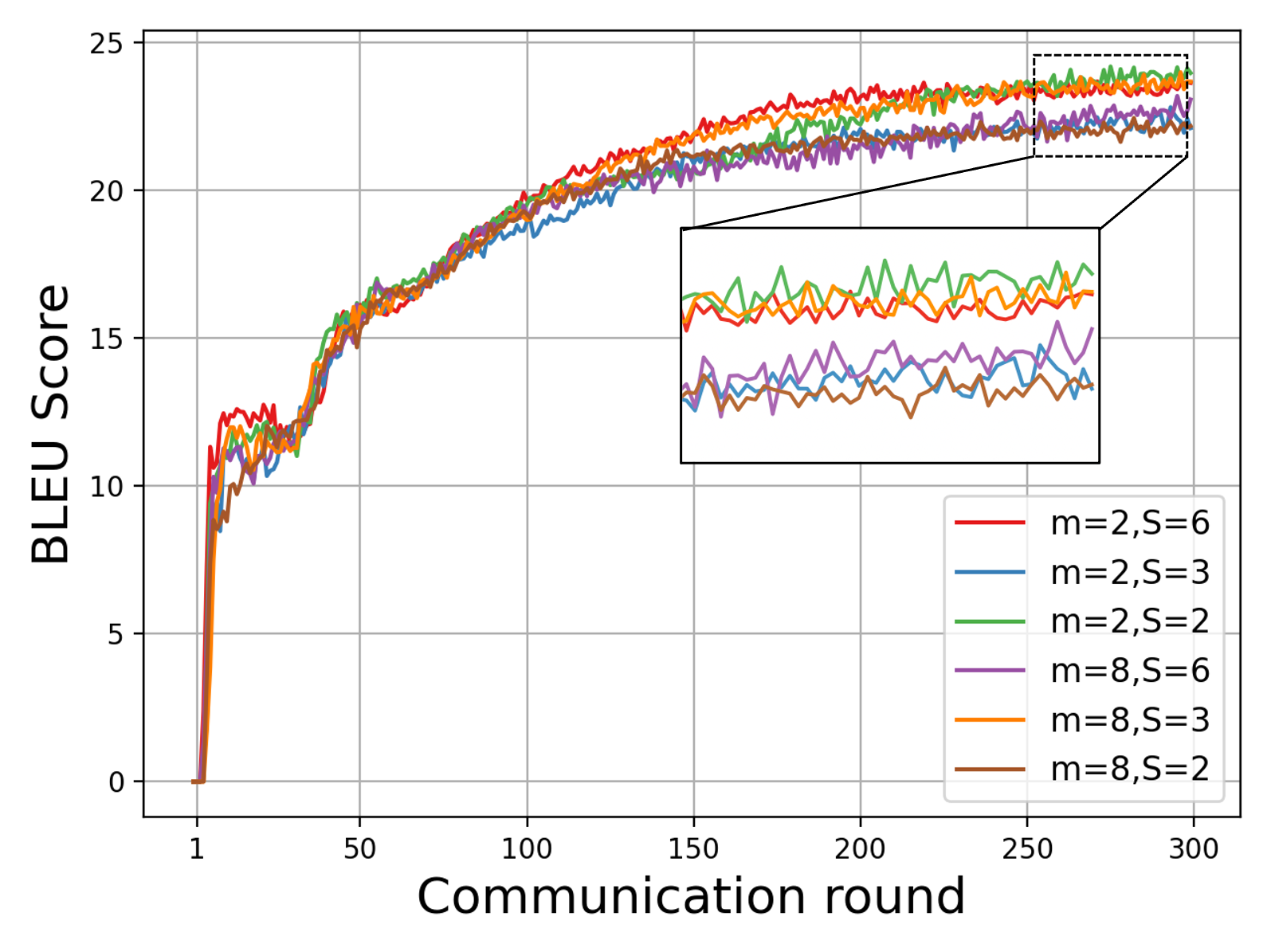}
	\caption{Learning performance of machine translation with different number of segment and micro-batch.}
	\label{fig9}
\end{figure}

\begin{figure}[t!]
	\centering
	\includegraphics[width=2.5 in]{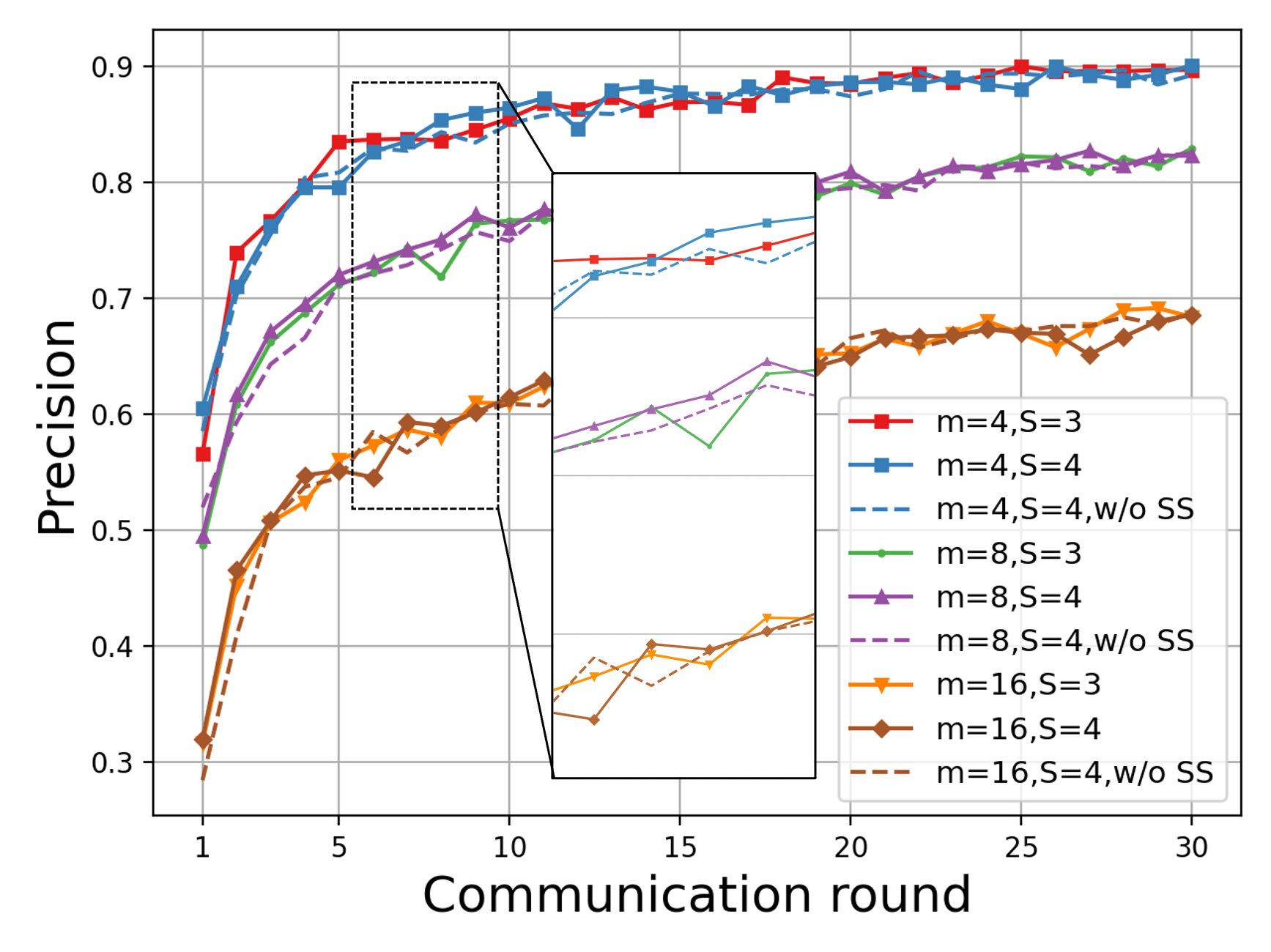}
	\caption{Learning performance of named entity recognition with different number of segment and micro-batch.}
	\label{fig10}
\end{figure}

\begin{figure}[t!]
	\centering
	\includegraphics[width=2.5 in]{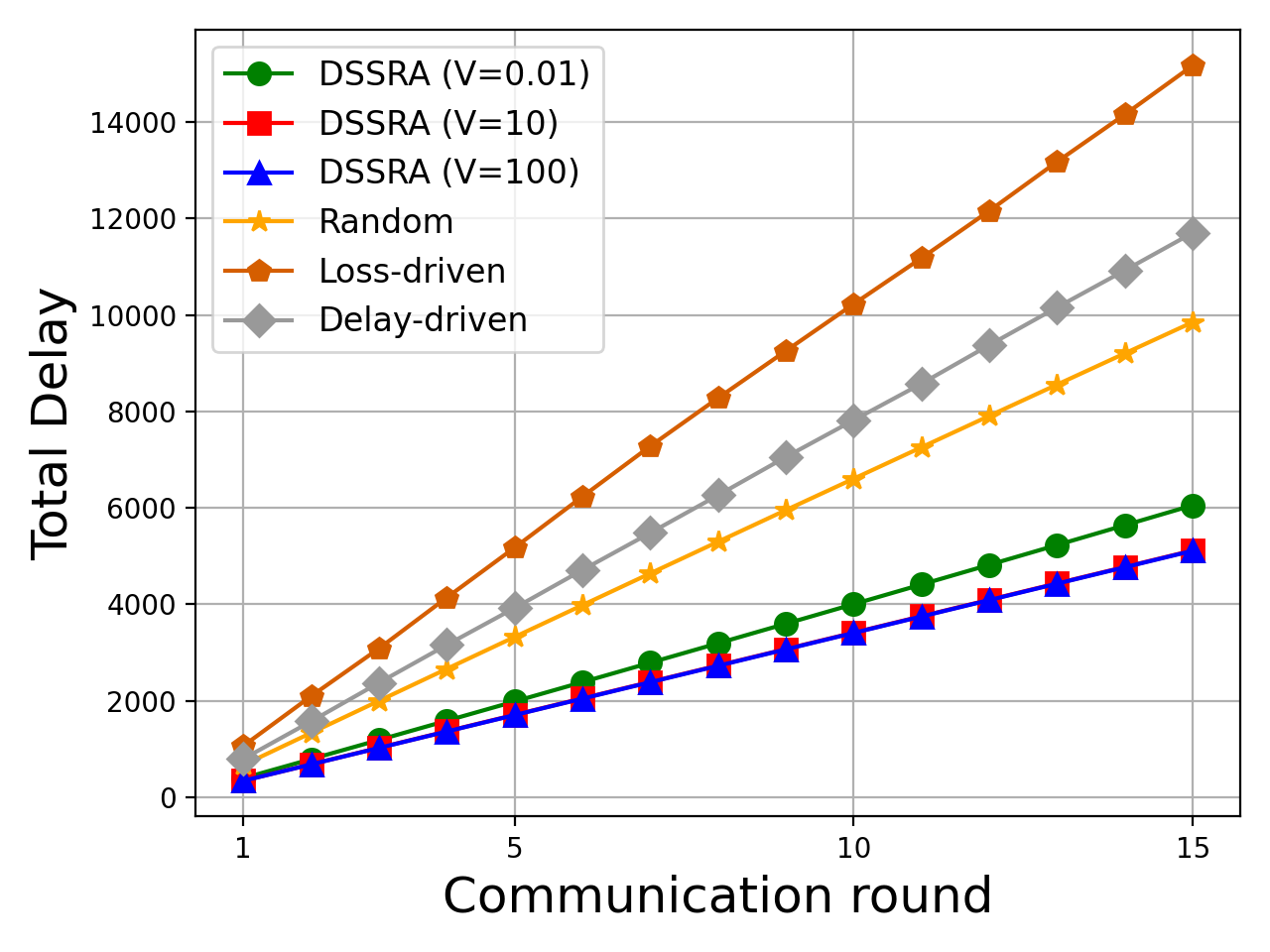}
	\caption{Total delay with different optimization schemes.}
	\label{fig11}
\end{figure}

Although FL does not incur additional D2D communication delay, it imposes high demands on the memory capacity of mobile devices and the availability of high-quality user data. 
Table~\ref{etab3} compares memory usage and data requirements across different methods. The CollaPipe framework offers greater flexibility by dynamically adjusting per-device memory usage based on the number of encoder blocks assigned during model partitioning, making it well-suited for resource-constrained edge environments.
Moreover, as training data is stored centrally in the control unit (CU), participating devices contribute only computational resources without needing to share local data. This significantly reduces the number of data-providing users and alleviates device management overhead in mobile edge networks.

\subsection{Sensitivity Analysis}
We conduct sensitivity analysis on the proposed DSSRA algorithm to illustrate the impact of joint optimization on system performance.

Fig. \ref{fig9} shows the inference performance with different numbers of segments $S$ and micro-batches $m$. For the Transformer model, the optimal configuration determined by the DSSRA algorithm is $S=2$ and $m=2$, under which the best learning performance is achieved. 
As shown in Fig.~\ref{fig10}, for the BERT model, disabling the segment scheduling module results in a slight reduction in inference performance. More notably, reducing the value of $m$ improves convergence performance, indicating a trade-off between parallel granularity and learning efficiency.

Fig. \ref{fig11} presents that the proposed DSSRA algorithm achieves lower system latency compared to the baseline schemes, and its advantage becomes increasingly pronounced as the number of communication rounds increases. Specifically, compared to the loss-only driven scheduling strategy, DDSRA with $V=0.01$ reduces the BERT training latency on the Toutiao dataset by 46.44\%. When compared to the delay-only driven scheduling, the latency reduction is 15.48\%, and against random scheduling, the DSSRA achieves a 7.12\% reduction. 
This improvement is attributed to our proposed dynamic optimization algorithm, which comprehensively accounts for the additional communication and computation overhead caused by frequent interactions at the Transformer block level. By effectively balancing intra-cluster and inter-cluster coordination, the algorithm enables efficient collaboration across heterogeneous networks.

\section{Conclusion}\label{s7}
This paper has presented CollaPipe, a novel hybrid parallel learning framework for collaborative LLM training. In the proposed approach, the encoder of the LLM is partitioned into $S$ segments in a TEB-wise manner and distributed across mobile devices, where the segments are trained using pipeline parallelism. After encoder training, each cluster’s CU communicates with the edge server to perform federated aggregation. 
To address challenges posed by heterogeneous and dynamic edge environments, 
%we introduced a selective scheduling and resource allocation strategy for each federated communication round, with the goal of minimizing end-to-end system latency.
we have formulated a bidirectional optimization framework that achieves a dynamic balance between learning performance and communication efficiency through a selective scheduling and resource allocation strategy. We have derived the closed-form convergence bound, establishing the relationship between system performance, the number of segments, and transmission power. To ensure long-term system stability, we have further integrated Lyapunov optimization with convergence analysis of the learning process.
Extensive experiments have validated the effectiveness of the CollaPipe framework in enhancing intra-cluster collaboration and improving training efficiency. The DSSRA algorithm has demonstrated its ability to achieve near-optimal system latency under resource-constrained and heterogeneous conditions.

Our findings also highlight the critical role of the number of micro-batches in collaborative learning, particularly for models like BERT. Future work will explore batch-level or token-level scheduling and optimization strategies in conjunction with model segmentation, aiming to further enhancing self-collaboration and self-intelligence within distributed intelligent IoT systems.

\section*{Appendix A}
\section*{Proof of Lemma 1}
Proof: According to (\ref{argmin}), we can expand the formula as
\begin{equation}\label{proof1-1}
	\begin{split}
		&\quad \left\Vert \nabla F(\theta^{FM}) \right\Vert^2 \\
		&= \left\Vert \frac{1}{N}\sum_{n=1}^{N} \nabla F_n(\theta_n^{FM}) \right\Vert^2 \\
		&= \left\Vert \frac{1}{N}\sum_{n=1}^{N} \left( \frac{\partial F_n(\tilde{\theta}^{dec})}{\partial \theta_n^{enc}}, \frac{\partial F_n(\theta_n^{enc})}{\partial \theta^{dec}} \right) \right\Vert^2.
	\end{split}
\end{equation}

Then we derive the derivatives for $\{ \theta_n^{enc} \}_{n\in \mathcal{N}}$ and $\theta^{dec}$ separately to obtain
\begin{equation}\label{proof1-2}
	\begin{aligned}
		&\quad \left\Vert \nabla F(\theta^{FM}) \right\Vert^2 \\
		&= \frac{1}{N^2} \sum_{n=1}^{N} \left\Vert \frac{\partial F_n(\tilde{\theta}^{dec})}{\partial \theta_n^{enc}} \right\Vert^2 + \frac{1}{N^2} \left\Vert \sum_{n=1}^{N} \left( \frac{\partial F_n(\theta_n^{enc})}{\partial \theta^{dec}} \right) \right\Vert^2\\
		&= \frac{1}{N^2} \left( \sum_{n\in \mathcal{N}}\left\Vert \nabla_{\theta_n^{enc}} F_n(\theta_n^{enc};\tilde{\theta}^{dec}) \right\Vert^2+ \right. \\
		&\quad \quad \quad \quad \quad \quad \quad \quad \quad \left. \left\Vert \sum_{n\in \mathcal{N}}\nabla_{\theta^{dec}}F_n(\theta^{dec};\theta_n^{enc}) \right\Vert^2 \right)\\
		&\overset{(a)}{\le} \frac{1}{N^2} \sum_{n\in \mathcal{N}} \left\Vert \nabla_{\theta_n^{enc}} F_n(\theta_n^{enc};\tilde{\theta}^{dec}) \right\Vert^2 + \\
		& \quad \quad \quad \quad \quad \quad \quad \quad \quad \frac{1}{N} \sum_{n\in \mathcal{N}} \left\Vert \nabla_{\theta^{dec}}F_n(\theta^{dec};\theta_n^{enc}) \right\Vert^2, 
	\end{aligned}
\end{equation}
where (a) follows by using the inequality $\Vert \sum_{i=1}^{n} \mathbf{z}_i \Vert^2 \le n\sum_{i=1}^{n}\Vert \mathbf{z}_i \Vert^2$ for any vectors $z_i$ and any positive integer $n$ (using $n=N$ in (a)).

The encoder is further partitioned into segments, and the parameters of each segment are split and substituted into
\begin{equation}
\left\Vert \nabla_{\theta_n^{enc}} F_n \right\Vert^2=\sum_{s=1}^{S}\left\Vert \nabla_{\theta_n^s} F_n \right\Vert^2,
\end{equation}
and we gain
\begin{equation}\label{proof1-3}
	\begin{split}
		&\quad \left\Vert \nabla_{\theta_n^{enc}} F_n \right\Vert^2 \\
		&\overset{(a)}{=} \left\Vert \frac{\partial F_n(\theta_n^{enc})}{\partial \theta_n^1} \right\Vert^2 + \left\Vert  \frac{\partial F_n(\theta_n^{enc})}{\partial \theta_n^2} \right\Vert^2 + \cdots + \left\Vert  \frac{\partial F_n(\theta_n^{enc})}{\partial \theta_n^S} \right\Vert^2  \\
		&\triangleq \left\Vert \frac{1}{\delta_1}\sum_{l=1}^{\delta_1} g_n^{1,l} \right\Vert^2 + \left\Vert  \frac{1}{\delta_2}\sum_{l=1}^{\delta_2}g_n^{2,l} \right\Vert^2 + \cdots + \left\Vert \frac{1}{\delta_S}\sum_{l=1}^{\delta_S}g_n^{S,l} \right\Vert^2 \\
		&\overset{(b)}{\le} \frac{S}{L} \sum_{l=1}^{L/S} \left( \left\Vert g_n^{1,l} \right\Vert^2 + \left\Vert g_n^{2,l} \right\Vert^2 + \cdots + \left\Vert g_n^{S,l} \right\Vert^2 \right),
	\end{split}
\end{equation}
where (a) follows from Eqs. (\ref{g-a}) and (\ref{g-c}); (b) follows by using the inequality $\Vert \sum_{i=1}^{n} \mathbf{z}_i \Vert^2 \le n\sum_{i=1}^{n}\Vert \mathbf{z}_i \Vert^2$ (using $n=\delta_i$ for (b)), $\delta_i$ indicates the average number of TEBs in a segment 
\footnote{To facilitate the proof, the same number of TEBs per segment is assumed only in the proof. The segment scheduling module in the CollaPipe framework allocates different numbers of TEBs according to different device capacities.}, so we remove $\delta_i$ by $\delta_i=L/S$.

Combing (\ref{proof1-2}) and (\ref{proof1-3}), (\ref{proof1-1}) can be upper bounded by
\begin{equation}
	\begin{split}
		&\quad \Vert \nabla F(\theta^{FM}) \Vert^2 \\
		&\le \frac{1}{N^2} \sum_{n\in \mathcal{N}}\left\Vert \frac{1}{\delta_i}\sum_{l=1}^{\delta_i}\sum_{s=1}^{S}\Vert g_n^{s,l} \Vert^2 \right\Vert^2 + \frac{1}{N}\sum_{n\in \mathcal{N}}\Vert \nabla_{\theta^{dec}}F_n \Vert^2 \\
		&\overset{(a)}{\le} \frac{S^2}{N^2L} \sum_{n\in \mathcal{N}} \sum_{l=1}^{L/S} \sum_{s=1}^{S} \left\Vert g_n^{s,l} \right\Vert^2 + \frac{1}{N}\sum_{n\in \mathcal{N}}\Vert \nabla_{\theta^{dec}}F_n \Vert^2,
	\end{split}
\end{equation}
where (a) follows by using the inequality $\Vert \sum_{i=1}^{n} \mathbf{z}_i \Vert^2 \le n\sum_{i=1}^{n}\Vert \mathbf{z}_i \Vert^2$ (using $n=S$ for (a)) and $\delta_i=L/S$.

The proof is completed.

\section*{Appendix B}
\section*{Proof of Theorem 1}
Before proceeding with the proof, recall that $\hat{\theta}$ represents the gradient after transmission through the interference channel, and $\widetilde{\theta}$ model parameters after global update.

According to Lemma 1, given the fixed decoder parameters, we can derive the upper bound on convergence related to the encoder's segment. In each communication round $t$ for FL, the global LLM update follows:
\begin{equation}\label{pf36}
	\widetilde{\theta}(t+1)=\hat{\theta}(t)- \frac{\eta}{N}\sum_{n=1 }^{N}\hat{g}_n(t),
\end{equation}
where
\begin{equation}
	\hat{g}_n(t)=\sum_{s=1}^{S}g^s_n(t) + \Delta g_n(t).
\end{equation}

According to Lemma 2, the channel interference results in the gradient error, we have 
\begin{equation}\label{proof2_error}
\mathbb{E} \Vert \Delta g_n(t) \Vert^2 \le \beta \epsilon(p_n(t)).
\end{equation}

To assist the derivation, we define a virtual model $v(t+1)$ that assumes no interference in the transmission signals between the mobile device and BS, that is, there is almost no error before and after data transmission, which is 
\begin{equation}\label{proof2-v}
	v_n(t+1) = \widetilde{\theta}(t) - \frac{\eta}{N}\sum_{n=1}^{N}g_n(t).
\end{equation}

Utilizing $\beta$-Smoothness, we can recursively obtain
\begin{equation}\label{proof2-Fv}
\begin{split}
	F(v(t+1)) &\le F(\theta(t)) - \eta\Vert\nabla F(\theta(t))\Vert^2 \\
	&\quad \quad \quad \quad + \frac{\beta\eta^2}{2N^2}\sum_{n=1}^{N}\Vert g_n(t)\Vert^2.
\end{split}
\end{equation}

The convergence upper bound of a full LLM can be decoupled into two modules: the encoder and the decoder. Next, we analyze the convergence with split learning of the encoder part. According to Lemma 1, the encoder gradient variance is split into $S$ segments:
\begin{equation}
\mathbb{E} \Vert g^{enc}_n(t)\Vert^2 \le \frac{S}{\delta_i} \sum_{s=1}^{S} \mathbb{E}\Vert g^s_n\Vert^2 \le \frac{S}{\delta_i}\phi^2.
\end{equation}

The overall gradient variance is:
\begin{equation}\label{proof2_all_gra}
\mathbb{E}\Vert \nabla F(\theta)\Vert^2\le \frac{S}{\delta_i}\phi^2+\frac{1}{N}\phi^2.
\end{equation}

We rewrite (\ref{proof2-Fv}) in the form of $F-F^{\ast}$ and take its expectation:
\begin{equation}\label{proof2_v-gradient}
\begin{split}
&\quad \mathbb{E} [F(v(t+1))-F^{\ast}] \\
&\le \mathbb{E} [F(\theta(t+1))-F^{\ast}]-\eta\Vert\nabla F(\theta(t))\Vert^2 \\
& \quad + \frac{\beta\eta^2}{2N^2}\sum_{n=1}^{N}\Vert g_n(t)\Vert^2 \\
&\overset{(a)}{\le} \mathbb{E}[F(\theta(t+1))-F^{\ast}] - 2 \eta \xi \mathbb{E}[F(\theta(t+1))-F^{\ast}] \\
&\quad + \frac{\beta\eta^2}{2N^2}\sum_{n=1}^{N}\Vert g_n(t)\Vert^2\\
&\overset{(b)}{\le} (1- 2 \eta \xi )\mathbb{E}[F(\theta(t+1))-F^{\ast}] + \frac{\beta \eta^2}{2N} \left(\frac{S}{\delta_i}+1\right) \phi^2,
\end{split}
\end{equation}
where (a) follows Assumption 5, substituting the PL condition into the second term; (b) follows (\ref{proof2_all_gra}) and combines like terms.

Subtracting (\ref{pf36}) from (\ref{proof2-v}), we obtain the gap between the actual parameter update and the virtual queue:
\begin{equation}
\theta(t+1)-v(t+1)= - \frac{\eta}{N}\sum_{n=1}^{N}\Delta g_n(t).
\end{equation}

According to Assumption 1, taking the expectation of the above function, the second-order term of $\beta$-smoothness is transformed into:
\begin{equation}\label{proof-59}
\begin{split}
\mathbb{E} \Vert \theta(t+1)-v(t+1)\Vert^2 &= \frac{\eta^2}{N^2}\mathbb{E}\Vert \sum_{n=1}^{N}\Delta g_n(t)\Vert^2 \\
& \overset{(a)}{\le} \frac{\eta^2\beta}{N}\epsilon(p_n(t)),
\end{split}
\end{equation}
where (a) follows Assumption 1 and Lemma 2.

Similarly, we rewrite (\ref{proof-59}) in the form of $F-F^{\ast}$ and take its expectation: 
\begin{equation}\label{proof2-adderro}
\mathbb{E} [F(\theta(t+1))-F^{\ast}] \le \mathbb{E} [F(v(t+1))-F^{\ast}]+\frac{\beta \eta^2}{2N}\epsilon(p_n(t)).
\end{equation}

Combing (\ref{proof2_v-gradient}) and (\ref{proof2-adderro}), we have
\begin{equation}\label{proof2-com}
\begin{split}
&\quad \mathbb{E}[F(\theta(t+1))-F^{\ast}] \\
&\le (1- 2 \eta \xi )\mathbb{E}[F(\theta(t+1))-F^{\ast}] + \frac{\beta \eta^2}{2N} \left(\frac{S}{\delta_i}+1\right) \phi^2 \\
&\quad \quad + \frac{\beta \eta^2}{2N}\epsilon(p_n(t)).
\end{split}
\end{equation}

Let $\sigma(t) = 2\eta\xi-\frac{\beta \eta^2}{2}\left(\frac{S}{N\delta_i}+\frac{1}{N}\right)$, $\epsilon(p_n) = \frac{C}{p_nh_n+I_i}$, then we expand (\ref{proof2-com}) from $t=0$ to $T-1$:
\begin{equation}\label{proof-62}
	\begin{split}
	&\quad F\left(\theta(T)\right) - F\left(\theta^{*}\right) \\
	&\le \left( \prod_{t=0}^{T-1}(1-2\sigma(t)) \right)F(\theta(0))-F(\theta^{*}) \\
	&\quad + \sum_{t=0}^{T-1}\prod_{j=t+1}^{T-1}\left(1-\sigma(j)\right)\left[\frac{\beta\eta^2 \phi^2}{N}\left(\frac{S^2}{L}+1\right)+\frac{\eta}{N}\epsilon(p_n(t))\right].
\end{split}
\end{equation}

According to (\ref{proof-62}), $\sigma(t)$ controls the degree of convergence in each round. As long as $\sigma(t)> 0$, the system can be guaranteed to converge. Let $\sigma(t)> 0$ and solve the system convergence condition:
$$
\eta < \frac{4\xi NL}{\beta(S^2+L)}.
$$

The proof is completed.

\bibliographystyle{IEEEtran.bst}
\bibliography{deviceAI.bib}

%\newpage
 
\vspace{5pt}

\vfill

\end{document}